\documentclass[aps,prb,twocolumn,showpacs,amsmath,amssymb,superscriptaddress,letterpaper]{revtex4}
\usepackage[pdftex]{graphicx} 
\usepackage{dcolumn} 
\usepackage{bm} 
\usepackage{relsize}
\usepackage{amsmath}
\usepackage{color}
\usepackage{psfrag}
\usepackage{epstopdf}
\usepackage{tocloft}
\usepackage[usenames,dvipsnames]{xcolor}
\definecolor{olive}{rgb}{0,0.6,0.4}
\begin{document}

\title{Spin-dependent optical properties in strained silicon and germanium}
\author{Pengke Li}
\affiliation{Department of Electrical and Computer Engineering, University of Rochester, Rochester, New York, 14627}
\author{Dhara Trivedi}
\affiliation{Department of Physics and Astronomy, University of Rochester, Rochester, New York, 14627}
\author{Hanan Dery}
\affiliation{Department of Electrical and Computer Engineering, University of Rochester, Rochester, New York, 14627}
\affiliation{Department of Physics and Astronomy, University of Rochester, Rochester, New York, 14627}

\begin{abstract}
We present a comprehensive theory of the circularly polarized luminescence and its dependence on strain in spin-polarized Si and Ge. Symmetries of wavefunctions and interactions are used to derive concise ratios between intensities of the right and left circularly polarized luminescence for each of the dominant phonon-assisted optical transitions. These ratios are then used to explain the circular polarization degrees of the luminescence peaks in the spectra of biaxially-strained Si and Ge, and of relaxed ${\rm{Si}}_{1-x}{\rm{Ge}}_{x}$ alloys. The spectra are numerically calculated by a combination of an empirical pseudopotential method, an adiabatic bond-charge model and a rigid-ion model.
\end{abstract}
\maketitle

\section{Introduction}

The conservation of angular momentum during the interaction of radiation with matter allows one to study the angular momentum of charge carriers from the state of light polarization.\cite{Zutic_RMP04,Qing_PRL11,OO_84} In semiconductors, the circular polarization degree is largely set by the effect of spin-orbit coupling on the angular-momentum quantum numbers in the valence band.\cite{OO_84} One can then use strain as a valuable experimental knob to regulate and validate the relation between the measured circular polarization degree and the spin polarization of carriers.\cite{OO_84,Crooker_PRL05} The applied strain lifts the energy degeneracy in the edge of the valence band, and it controls not only the energy spacings but also the mixing between hole species. Knowing the strain-induced hole mixing in each of the split valence bands, one can ascertain the spin polarization from the measured circular polarization degree. To date, this technique has been widely used in direct-gap semiconductors where optical transitions are carried out straightforwardly.\cite{Zutic_RMP04,OO_84}

In indirect absorption edge semiconductors, free-carrier optical transitions are more intricate since they involve both radiation-matter and electron-phonon interactions.\cite{Lax_PR61,Nakayama_JJAP80,Thewalt_chapter,Wagner_PRB84} The pho\textbf{t}on has negligible momentum in comparison to charge carriers, and therefore the emission or absorption of a pho\textbf{n}on is needed to conserve the crystal momentum during electron-hole radiative recombination (between edges of the conduction and valence bands).  Extracting information on the spin polarization of electrons from the circular polarization degree of phonon-assisted optical transitions was recently established in the case of unstrained silicon.\cite{Li_PRL10,Cheng_PRB11} This interpretation allows one to quantify the spin polarization of electrons in experiments where strain is not a controllable parameter.\cite{Jonker_NaturePhysics07,Grenet_APL09,Kioseoglou_APL09,Li_APL09,Jansen_PRB10,Pezzoli_PRL12}

In this paper we study the effect of strain on the luminescence in spin-polarized Si and Ge. We derive concise ratios between intensities of the right and left circularly polarized luminescence for each of the optical transitions in these materials. These ratios depend on the phonon type, hole mixing, and valley position within the multivalley conduction band. The derived ratios are then used to interpret numerically calculated spectra in biaxially-strained Si and Ge as well as in relaxed ${\rm{Si}}_{1-x}{\rm{Ge}}_{x}$ alloys. Given the predictions for long spin lifetimes in strained
Si and Ge,\cite{Dery_APL10,Tang_PRB12,Song_PRB12,Li_PRB12} the findings of this work render optical transitions a viable tool in studying the spin dynamics when straining these materials. The dependence of spin relaxation on strain can be determined from the change in the circular polarization degree when the spin lifetime is comparable or longer than the electron recombination timescale. In this limit, the circular polarization degree is changed from zero to the maximal attainable theoretical value (that will be presented in this work).

This paper is organized as follows. In Sec.~\ref{sec:frame} we review the theory of free-carrier optical transitions in indirect absorption edge semiconductors. After summarizing known spin-independent results in Ge and Si, we present a detailed procedure to derive spin-dependent selection rules. In Sec.~\ref{sec:strain} we derive expressions for the  circular polarization degree as a function of the strain amplitude and type. This section also includes a brief review of strain effects on the energy band structure.  In Sec.~\ref{sec:spectra} we confirm the analytical findings by independent numerical calculations that combine an empirical pseudopotential method, an adiabatic bond-charge model and a rigid-ion model. This numerical procedure is used to calculate the polarized spectra in biaxially-strained Si and Ge, as well as in relaxed ${\rm{Si}}_{1-x}{\rm{Ge}}_{x}$ alloys. Section~\ref{sec:sum} is a summary of central results. Appendix~\ref{KL} includes technical details of the strain-dependent Luttinger-Kohn Hamiltonian model, Appendix~\ref{phonon} deals with the moderate effect of strain on the phonon dispersion, and Appendix~\ref{OO} includes numerical results of optical orientation in unstrained Si and Ge.

\section{Background}\label{sec:frame}
The intensities of optical transitions in indirect absorption edge semiconductors are quantified by using second-order perturbation theory.\cite{Yu_Cardona_Book} The radiation-matter and electron-phonon interactions are denoted by  $H_{\hat{\mathbf{e}}}$ and $H_{\ell}$, respectively. $\hat{\mathbf{e}}$ is the light polarization vector and $\ell$ is the phonon mode: transverse-acoustic (TA), transverse-optical (TO), longitudinal-acoustic (LA) or longitudinal-optical (LO). The intensity of an optical transition is then found proportional to
\begin{equation}
I_{\hat{\mathbf{e}},\ell}^{i \rightarrow f} \!\propto \! \left| \sum_n \! \frac{\langle f|H_{\hat{\mathbf{e}}}|n\rangle \langle n|H_{\ell}|i\rangle}{E_i-E_n-\hbar\omega_\ell}\!+\! \frac{\langle f|H_{\ell}|n\rangle\langle n| H_{\hat{\mathbf{e}}} |i\rangle }{E_i-E_n \mp \hbar\omega_{0}}\right|^2\!\!,
\label{eq:intensity}
\end{equation}
where the angular frequency of the photon is $\omega_0$, and the $\mp$ sign is associated with luminescence ($-$) or absorption ($+$). $\omega_\ell$ is the angular frequency of the phonon emitted during the process. The initial and final states are given by $|i\rangle$ and $|f\rangle$, respectively. In luminescence and optical orientation experiments these states are typically taken from the edges of the conduction and valence bands. The overall energy conservation implies that $E_f-E_i = \hbar(\omega_0 \pm \omega_{\ell}$). The sum in Eq.~(\ref{eq:intensity}) is over intermediate states ($|n\rangle$). The translation symmetry renders crystal momentum conservation in each of the radiation-matter and electron-phonon interactions (i.e., during virtual transitions to and from intermediate states). The interference between the first and second terms in Eq.~(\ref{eq:intensity}) plays an important role in setting the amplitude of phonon-assisted optical transitions in Si.\cite{Lax_PR61,Glembocki_PRL82,Glembocki_PRB82,Bednarek_PRL82,Li_PRL10}

\subsection{Spin-independent properties} \label{subsec:independent}
\begin{figure}
\centering
\includegraphics[width=8.5 cm]{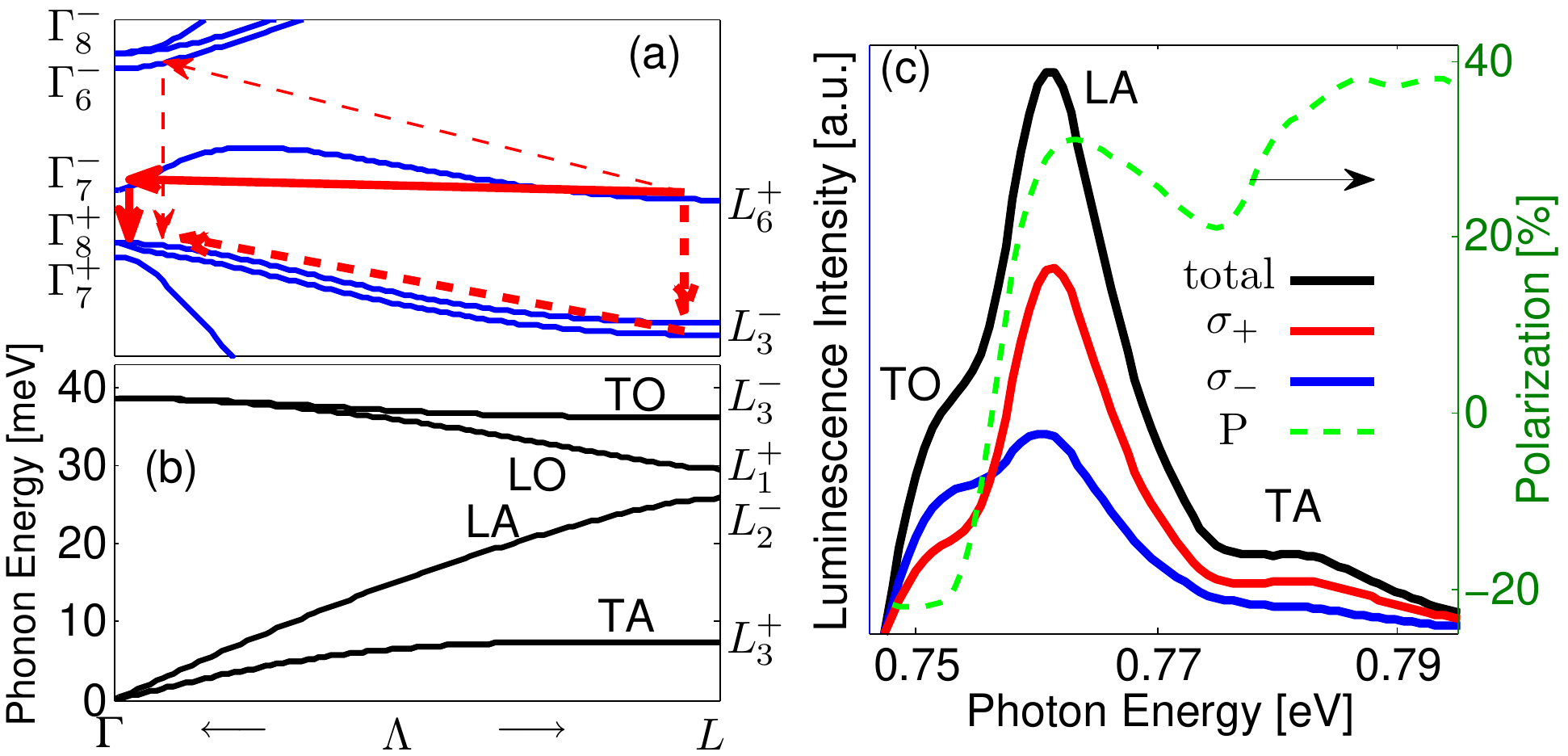}
\caption{(color online) (a) Band structure of Ge along the $\Gamma-\Lambda-L$ symmetry axis. The conduction (valence) edge state is indicated by $L_6^+$ ($\Gamma_8^+$). Arrows represent virtual paths for phonon-assisted optical transitions via pertinent intermediate states. The path via $\Gamma_{7}^-$ ($L_3^-$) is shown in bold solid (dash) lines to represent its dominant role in setting the intensity of LA (TO) phonon-assisted optical transitions.\cite{footnote_L3}  (b) Phonon dispersion along the $\Gamma-\Lambda-L$ symmetry axis. (c) Calculated polarized luminescence due to radiative recombination of spin-up electrons in unstrained Ge at 77~K.}\label{fig:Geband}
\end{figure}

To understand the salient features of the luminescence or absorption spectra, one has to inspect the energy band structure and the phonon dispersion curves. As an example, in Fig.~\ref{fig:Geband} we show this information for the case of unstrained Ge between the $\Gamma$ and $L$ points. The band structure in Fig.~\ref{fig:Geband}(a) is from an empirical pseudopotential method,\cite{Cheli_PRB76} and the phonon dispersion in Fig.~\ref{fig:Geband}(b) is from an adiabatic bond-charge model.\cite{Weber_PRB77} The procedure to calculate the luminescence spectrum in Fig.~\ref{fig:Geband}(c) will be explained in Sec.~\ref{sec:spectra}. The top edge of the valence band is two-band degenerate and located in the zone center [indicated by $\Gamma_8^+$ in Fig.~\ref{fig:Geband}(a)]. The bottom edge of the nondegenerate conduction band is indicated by $L_6^+$ and located in four equivalent zone-edge $L$~points ($[\pi,\pi,\pi]/a$, $[-\pi,\pi,\pi]/a$, $[\pi,-\pi,\pi]/a$ and $[\pi,\pi,-\pi]/a$ where $a$ is the lattice constant). Below we explain the origin for the three edge-to-edge luminescence features of the black curve in Fig.~\ref{fig:Geband}(c) as well as the origin for the difference in their amplitudes.

The luminescence process is schematically described by combination of arrow pairs in Fig.~\ref{fig:Geband}(a). The vertical arrows correspond to the radiation-matter interaction and the arrows between $L$ and $\Gamma$ correspond to the interaction of electrons with zone-edge phonons near the $L$ point. The transition from initial to final states can proceed via several virtual paths. From inspection, however, the energy proximity between the $L_6^+$ and $\Gamma_7^-$ states of the conduction band is evident ($\sim$140~meV). The phonon-assisted optical transitions in Ge are governed by the $\Gamma_7^-$ intermediate state since it comes with the smallest `penalty' [denominator in Eq.~(\ref{eq:intensity})]. Using group theory, Lax and Hopfield have shown that only LA modes can be involved in the $L_6^+ \rightarrow \Gamma_7^- \rightarrow \Gamma_8^+ $ optical transition [solid arrows in Fig.~\ref{fig:Geband}(a)].\cite{Lax_PR61}

In addition to the relatively strong LA spectral peak, the spectrum of unstrained Ge involves TO and TA spectral features.\cite{Haynes_JPCS59} When considering `distant' intermediate states in Ge (the $\Gamma_6^-$, $\Gamma_8^-$ and $L_3^-$), the TO mode is also symmetry-allowed in edge-to-edge optical transitions,\cite{Lax_PR61} depicted by dashed arrows in Fig.~\ref{fig:Geband}(a). However, as elucidated by Glembocki and Pollak,\cite{Glembocki_PRL82,Glembocki_PRB82} the intensity of transition via $L_3^-$ is much stronger than via  $\Gamma_6^-$ and $\Gamma_8^-$.\cite{footnote_L3} Specifically, the matrix element from electron-phonon interaction between $L_3^-$ and  $\Gamma_8^+$  is $\sim$5 times larger than that between $L_6^+$ and $\{\Gamma_6^-$, $\Gamma_8^-\}$. As shown in Fig.~\ref{fig:Geband}(c) and supported by luminescence experiments,\cite{Haynes_JPCS59} one can also find a weak spectral peak from TA modes. In spite of the fact that edge-to-edge optical transitions with the TA mode are forbidden, higher order transitions between states near $L_6^+$ and $\Gamma_7^-$ do not vanish completely. The luminescence peak of the TA mode is visible since this phonon has exponentially larger population compared with that of other phonons. This property can be inferred by inspection of the $L$$\,$-point phonon energies [Fig.~\ref{fig:Geband}(b)].

In silicon, the absence of a single close-by intermediate state renders the intensity of optical transitions strongly susceptible to quantum-mechanical interference.  The six bottom edges of the multivalley and nondegenerate conduction band in unstrained silicon are positioned $\sim$0.85$\times$2$\pi/a$ away from the zone center on the $\Gamma - \Delta - X$ symmetry axes. Invoking group theory selection rules, Lax and Hopfield deduced a dominant involvement of transverse phonons whereas longitudinal phonons suffer from destructive interference between valence and conduction intermediate states [first and second terms in Eq.~(\ref{eq:intensity})].\cite{Lax_PR61}   In experiments, however, only the TO spectral peak is relatively strong while the TA spectral peak is closer in amplitude to that of the weak LO transition.\cite{Nakayama_JJAP80,Thewalt_chapter} This behavior is explained by a relatively weak interaction of electrons with zone-edge TA phonons [the $H_{\ell=TA}$ matrix element in Eq.~(\ref{eq:intensity})]. It stems from destructive interference between contributions of the two atoms in the unit cell (not to be confused with the interference between intermediate states).\cite{Glembocki_PRL82,Glembocki_PRB82,Bednarek_PRL82,Li_PRL10}

\subsection{Spin-dependent luminescence in Si and Ge} \label{subsec:circ_si}
To establish connection with the strain-dependent analysis in the next sections, we will consider the application of infinitesimal strain along $\hat{\mathbf{n}}$ and calculate the optical selection rules for light propagation along this direction. On the one hand, this procedure provides physical insight to the expected selection rules of strained crystals since it correctly classifies the zone-center eigenstates of the valence band. On the other hand, this procedure also allows us to extract the selection rules of unstrained crystals by neglecting the infinitesimally-induced energy splits in the edges of the conduction and valence bands. We mention that the spin-dependent optical selection rules in unstrained Si or Ge are independent of the choice of light propagation ($\hat{\mathbf{n}}$). This independence is noticed from the fact that in the absence of a symmetry-breaking mechanism, the angular-momentum conservation law is invariant to the choice of the spin-quantization axis.

For the general case of light propagation along $\hat{\mathbf{n}}$, the radiation operators of the left and right circular polarization vectors are
\begin{eqnarray}
{\sigma}_{\pm}&=&\frac{1}{\sqrt{2}}\left[\left(p_x\cos\theta\cos\phi+p_y\cos\theta\sin\phi+p_z\sin\theta\right)\right. \nonumber\\ && \left.\pm i\left(p_y\cos\phi-p_x\sin\phi\right)\right], \label{eq:light_n}
\end{eqnarray}
where $p_i$ are vector components of the momentum operator and $\theta$ ($\phi$) are the polar (azimuthal) angles of $\mathbf{n}$ with respect to the crystallographic axes. The degenerate spin states of conduction electrons in the edges of the valleys are replaced by,\cite{Song_PRB12}
\begin{eqnarray}
\!\!\!\!\!\!\!\!|\mathbf{K}_{j}, \Uparrow_\mathbf{n} \rangle &=&\cos\frac{\theta}{2}|\mathbf{K}_{j}, \Uparrow_z \rangle +\sin\frac{\theta}{2} e^{i\phi}|\mathbf{K}_{j}, \Downarrow_z \rangle, \,\,\,\,\,\, \nonumber \\
\!\!\!\!\!\!\!\! |\mathbf{K}_{j}, \Downarrow_\mathbf{n} \rangle &=&-\sin\frac{\theta}{2} e^{-i\phi}|\mathbf{K}_{j}, \Uparrow_z \rangle + \cos\frac{\theta}{2}|\mathbf{K}_{j}, \Downarrow_z \rangle, \,\,\,\,\,\, \label{eq:spin_up_n}
\end{eqnarray}
where $K_j$ is the wavevector in the center of the $j^{th}$ valley.

The classification of holes states is subtle and depends on the form of the strain tensor. Below we discuss two common cases where an infinitesimal stress is applied parallel to the $[001]$ and $[111]$ crystallographic axes. The former (latter) represents a case where the energy degeneracy between conduction-band valleys in Si (Ge) is lifted. These strain configurations also capture the important differences induced by diagonal and off-diagonal strain-tensor components. For example, the leading-order effect of [001]-strain is to couple light-hole and split-off states (both have $J_z=\pm 1/2$ components), while the heavy-hole states are left unaffected ($J_z=\pm 3/2$). On the other hand, of the two topmost valence bands that are split by [111]-strain, the one with heavier effective mass is described by combination of $\{+3/2, +1/2\}$ eigenstates of $J_{[111]}=\mathbf{J}\cdot(\hat{\mathbf{x}}+\hat{\mathbf{y}}+\hat{\mathbf{z}})/\sqrt{3}$, while the band with lighter effective mass is described by combination of $\{-3/2, -1/2\}$ eigenstates of $J_{[111]}$. Detailed calculations of these well-established results are given in Appendix~\ref{KL}.

\begin{table} \caption{
Representing basis functions of initial, final and intermediate states in the process of phonon-assisted luminescence of spin-up electrons in germanium. Capital letters denote even functions, $X=yz$, $Y=zx$ and $Z=xy$. See the band structure [Fig.~\ref{fig:Geband}(a)] for positions of these states.
}\label{Table:1}
\begin{center}
\renewcommand{\arraystretch}{1.8}
\begin{tabular}{c | c c}
\hline
initial state & \raisebox{.5ex}[0pt]{$L_{6,\frac{1}{2}}^+$} & $\frac{1}{\sqrt{3}}(X+Y+Z)\uparrow$ \\ \hline
 & \raisebox{.5ex}[0pt]{HH ($\Gamma_{8,``+\frac{3}{2}"}^{+}$)} & $\frac{1}{\sqrt{2}} (X+ i Y) \uparrow $ \\ \cline{2-3}
final states& \raisebox{-3ex}[0pt]{LH ($\Gamma_{8,``\pm\frac{1}{2}"}^{+}$)} & \raisebox{-.8ex}[0pt]{$\frac{1}{\sqrt{6}} [-(X- i Y) \uparrow - 2Z \downarrow]$,}\\
\raisebox{3ex}[0pt]{\,}&&\raisebox{.8ex}[0pt]{$\frac{1}{\sqrt{6}} [(X+ i Y) \downarrow - 2Z \uparrow]$} \\  \hline
  &
$\Gamma_{7}^-$    & $xyz\uparrow$ \\ \cline{2-3}
\raisebox{1ex}[0.pt]{intermediate} &$\Gamma_{6}^-+\Gamma_{8}^-$   & $x\uparrow $, $y\uparrow $, $z\uparrow $ \\ \cline{2-3}
\raisebox{3ex}[0pt]{states}& $L_{3}^-$ & $\frac{1}{\sqrt{2}}(x-y)\uparrow$, $\frac{1}{\sqrt{6}}(2z-x-y)\uparrow$\\ \hline
\end{tabular}
\end{center}
\end{table}

$\hat{\mathbf{n}} \parallel [001]$: The classification of the zone-center hole states in the presence of infinitesimal [001]-strain is similar to that of unstrained crystals (Appendix~\ref{KL}). To obtain the intensity ratios between right and left circularly polarized luminescence we calculate the relative amplitudes of radiation-matter and electron-phonon matrix elements by the use of basis functions. Using the previous analysis and the group theory notations in Fig.~\ref{fig:Geband}, the luminescence in Ge is employed as a case study example. Following Sec.~\ref{subsec:independent}, the conduction minima is represented by $L_6^+$, and we take the spin-up $L_{6,\frac{1}{2}}^+$ as the initial state. According to this choice, Table~\ref{Table:1} lists the representation basis functions of all states that take part in optical transitions. The notation of subscripts with quotation marks is to recall that the spin-orbit coupling is not with odd $p$-type orbitals but with even valence states. The basis functions of all possible intermediate states [$|n\rangle$ in Eq.~(\ref{eq:intensity})] are listed by the direct product of spin-\textit{up} and the basis functions of single-group irreducible representations. From inspection of Fig.~\ref{fig:Geband}(a), they are associated with $\Gamma$-point conduction states and with $L$-point valence states. In Table~\ref{Table:1}, we have neglected the spin mixing of intermediate states caused by the relatively weak spin-orbit interaction [reflected by the small energy split between $\Gamma_8^{-}$ and $\Gamma_6^{-}$, and between the $L_3^{-}$ states, see Fig.~\ref{fig:Geband}(a)]. However, since the summation in Eq.~(\ref{eq:intensity}) is complete, using the concise intermediate-state expressions of Table~\ref{Table:1} does not affect the final result of the analysis, while it simplifies the derivation.

\begin{table}[ht] \caption{Representing basis functions of zone edge phonons.}\label{Table:Phonon symmetry}
\begin{center}
\renewcommand{\arraystretch}{1.8}
\begin{tabular}{c|c}
Mode (symmetry)             &  Basis Function    \\ \hline
TO ($L_3^-$)                 & $\frac{1}{\sqrt{2}}(x-y)$,$\frac{1}{\sqrt{6}}(2z-x-y)$        \\ \hline
LO ($L_1^+$)                  &      1                \\ \hline
LA ($L_2^-$)                 &   $\frac{1}{\sqrt{3}}(x+y+z)$             \\ \hline
TA ($L_3^+$)            & $\frac{1}{\sqrt{2}}(x^2-y^2)$,$\frac{1}{\sqrt{6}}(2z^2-x^2-y^2)$             \\ \hline
\end{tabular}
\end{center}
\end{table}

\begin{figure}
\includegraphics[width=8.5 cm]{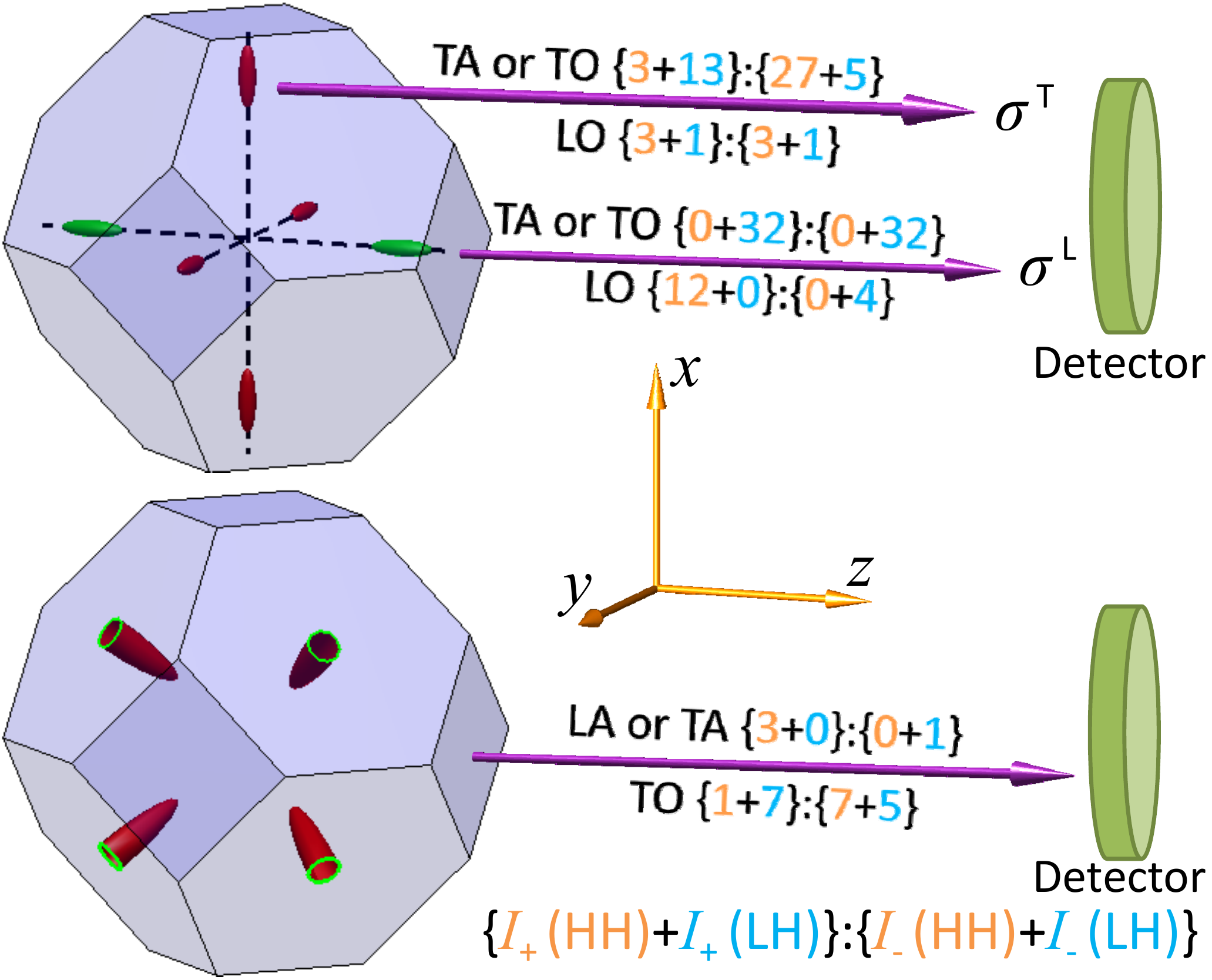}
\caption{(color online) Selection rules of phonon-assisted optical transitions in Si (upper panel) and Ge (lower panel). These rules pertain to transitions with electrons whose spins are oriented along the propagation direction of the detected light (+$z$ crystallographic axis). Next to each of the indicated phonon modes the ratio $\{$\textcolor{orange}{\textbf{A}}+\textcolor{cyan}{\textbf{B}}:\textcolor{orange}{\textbf{C}}+\textcolor{cyan}{\textbf{D}}$\}$ denotes the intensity ratio between right and left circularly polarized luminescence. Contributions from transitions with heavy (light) holes are denoted by \textcolor{orange}{\textbf{A}}:\textcolor{orange}{\textbf{C}} (\textcolor{cyan}{\textbf{B}}:\textcolor{cyan}{\textbf{D}}).}
\label{fig:sr_Ge_Si}
\end{figure}

\begin{figure}
\includegraphics[width=8.5cm]{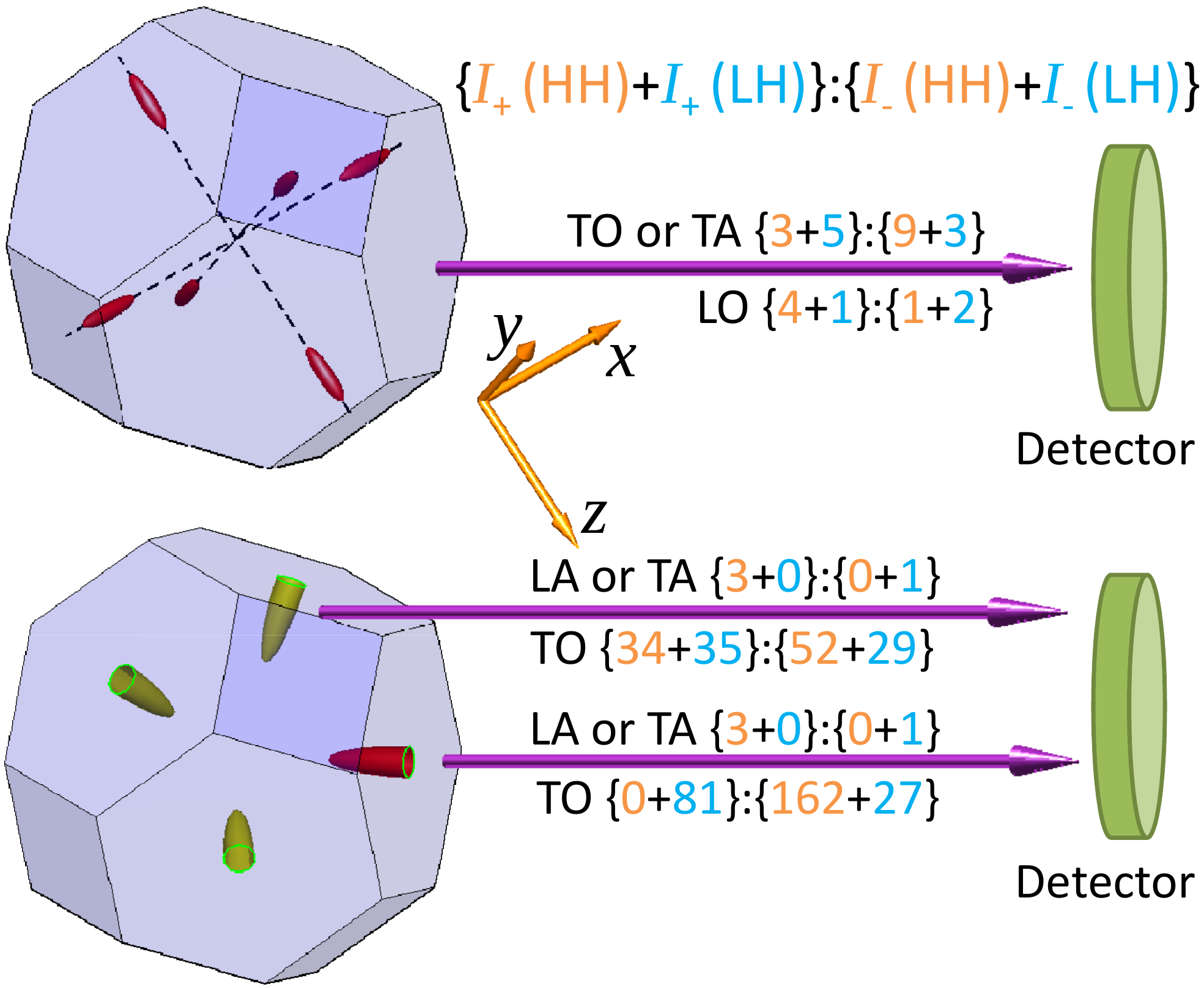}
\caption{(color online) Selection rules for light propagation along the [111] crystallographic axis.  Similar to the case in Fig.~\ref{fig:sr_Ge_Si}, numbers denote relative intensities of right and left circularly polarized luminescence of a certain phonon mode. These numbers should not be used to infer the relative intensities of different phonon modes.}
\label{fig:sr_Ge_Si_111}
\end{figure}

The \textit{relative} amplitudes of electron-phonon matrix elements can be calculated by considering the symmetries of zone-edge phonons. Table~\ref{Table:Phonon symmetry} lists the representation basis functions of these phonons. Together with the radiation operators of the two circular polarization vectors, ${\sigma}_{\pm}=({p}_{x}\pm i{p}_{y})/\sqrt{2}$, the selection rules are established straightforwardly. For example,
\begin{eqnarray}
&& L_{6,\frac{1}{2}}^{+}   \stackrel{\rm{LA}}{\longrightarrow}\Gamma_{7,\frac{1}{2}}^{-}  \stackrel{\sigma_+}{\longrightarrow} \Gamma_{8,``\frac{3}{2}"}^{+}  \propto \nonumber \\ && \left \langle \Gamma_{8,``\frac{3}{2}"}^{+} |{\sigma}_{+}| \Gamma_{7,\frac{1}{2}}^{-}\right  \rangle \left \langle \Gamma_{7,\frac{1}{2}}^{-}|\frac{x+y+z}{\sqrt{3}}| L_{6,\frac{1}{2}}^{+}  \right \rangle  \nonumber = P_1P_2, \nonumber
\end{eqnarray}
where $P_{1}=\langle xyz|x|X \rangle$ and $P_{2}=\langle X|{p}_{x}|xyz \rangle$ (and similarly for their cyclic permutations). Figure~\ref{fig:sr_Ge_Si} summarizes the resulting selection rules in Si and Ge. For each of the phonon modes, we list the intensity ratios of the right and left circularly polarized light due to contributions of heavy-hole (HH) and light-hole (LH) final states.\cite{footnote_nostrain} As shown in Fig.~\ref{fig:sr_Ge_Si}, light propagation along the $z$-crystallographic axis breaks the symmetry between valleys in Si but not in Ge. Therefore, the selection rules in Si are different for transverse or longitudinal valleys with respect to the propagation direction. $\sigma^T$ and $\sigma^L$ denote the respective emitted circularly polarized light.

$\hat{\mathbf{n}} \parallel [111]$: We use zone-center hole states in the presence of infinitesimal [111]-strain (Appendix A), and repeat the previous analysis. The resulting selection rules are shown in Fig.~\ref{fig:sr_Ge_Si_111}, where in Ge these rules can be different for the $L_{111}$ valley and for the other valleys. For each of the phonon modes, we list the relative intensity ratios of the right and left circularly polarized light due to contributions of HH and LH states. In this strain configuration, the hole states are no longer associated with pure $\pm3/2$ or $\pm1/2$ magnetic quantum numbers. Nonetheless, the selection rules are invariant to the choice of $\mathbf{n}$ when summing the contributions from both types of holes and from all valleys. This summation is relevant in unstrained crystals leading in the case of bulk Si to intensity ratios of $\sigma_{+}$$\,$$:$$\,$$\sigma_{-}$$\,$=$\,$$2$\,$:$\,$3$ for either the TO or TA peaks, and 5$\,$:$\,$3 for the LO peak. In unstrained bulk Ge, this summation leads to a 3$\,$:$\,$1 intensity ratio for either the LA or TA peaks, and 2$\,$:$\,$3 for the TO peak.

We conclude this section with mentioning two important aspects. First, the `direct-gap type' 3$\,$:$\,$1 intensity ratio for either the LA or TA peaks in Ge is independent of the valley position. The reason is that their dominant optical transitions involve electron-phonon interaction between relatively pure spin states in the lowermost conduction band. The second aspect relates to the energy proximity of TO and LO phonons in Si (both are $\sim$60~meV with a spacing of $\sim$4~meV). Having separate TO and LO spectral peaks in Si is feasible only at very low temperatures and high-purity silicon.\cite{Nakayama_JJAP80,Thewalt_chapter} In all other cases the two spectral peaks merge,\cite{Wagner_PRB84} and one should recall that this single peak has a large (small) contribution from TO (LO) phonon-assisted optical transitions.  One should also consider the opposite sign of their circular polarization degrees: $-$20\% for TO and +25\% for LO in unstrained Si. Due to the opposite sign contributions, the overall circular polarization degree of this peak is expected to be lower than that of the less intense TA peak (in spite of the similar selection rules for TA and TO transitions). This physics elucidates the observed difference in the measured circular polarization degrees of the dominant and TA peaks in recent experiments.\cite{Jonker_NaturePhysics07,Kioseoglou_APL09,Li_APL09,Jansen_PRB10} Quantitative analysis of unstrained Si is found in Ref.~[\onlinecite{Li_PRL10}].

\section{Strain Effects} \label{sec:strain}

We study the effect of strain amplitude on the selection rules in Si and Ge. The results of this study will be used in Sec.~\ref{sec:spectra} to elucidate features of the spin-polarized spectra that cannot be explained with infinitesimal-strain selection rules (Figs.~\ref{fig:sr_Ge_Si}~and~\ref{fig:sr_Ge_Si_111}). Below we first describe the effect of strain on the energy bands, and then we derive selection rules in the presence of a finite biaxial strain. Appendix~\ref{phonon} includes a brief description of the strain-induced modification to the phonon dispersion curves.

\begin{figure} 
\includegraphics[width=8.5cm]{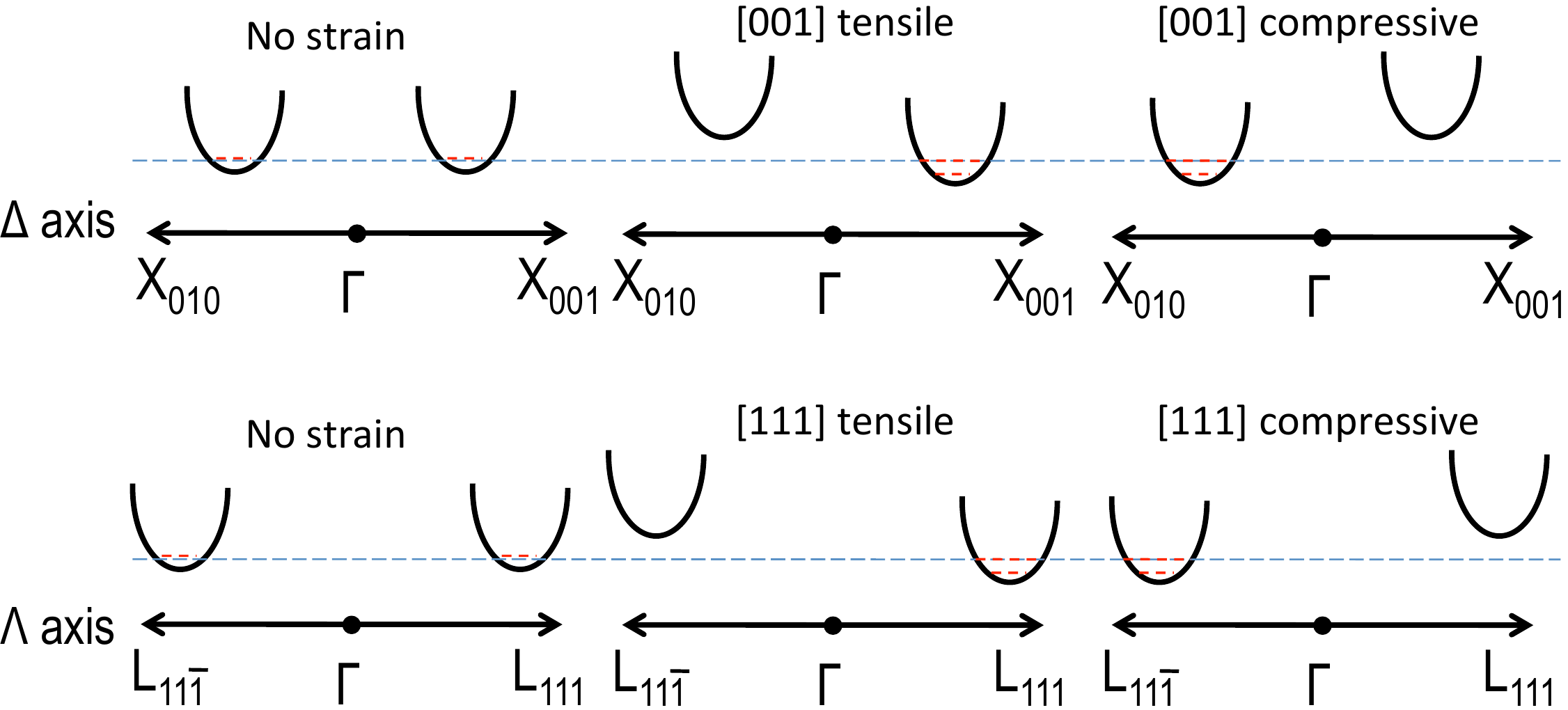}
\caption{Energy shift of conduction-band valleys under biaxial strain. Upper (lower) panels represent the case of Si (Ge) where low-energy valleys are centered on the $\Gamma$-$\Delta$-$X$ symmetry axes ($L$ points in the edges of the $\Lambda$ symmetry-axes). Contribution to the luminescence is governed by thermal electrons that populate the low-energy valley(s).} \label{fig:strainband}
\end{figure}

We start with the effect of strain on the multivalley conduction band. Figure~\ref{fig:strainband} shows diagrams of the energy shifts of $\Delta$ and $L$ valleys under [$001$] and [$111$] biaxial strain, respectively. Valleys shift up or down in energy depending on the angle between valley and strain axes and on whether the strain is compressive or tensile.\cite{Herring_PR56,Hensel_PR65,Laude_PRB71,Fischetti_JAP96,Sun_JAP07} Evidently, the energy split between valleys suppresses the intervalley electron scattering. As a result, the charge mobility is improved.\cite{Welser_IEEE94,Rim_IEEE00,Lee_JAP05} The spin lifetime is expected to increase more dramatically due to its strong dependence on intervalley processes.\cite{Cheng_PRL10,Li_PRL11,Li_PRL12,Dery_APL10,Tang_PRB12,Song_PRB12,Li_PRB12}

\begin{figure} 
\includegraphics[width=8.5cm]{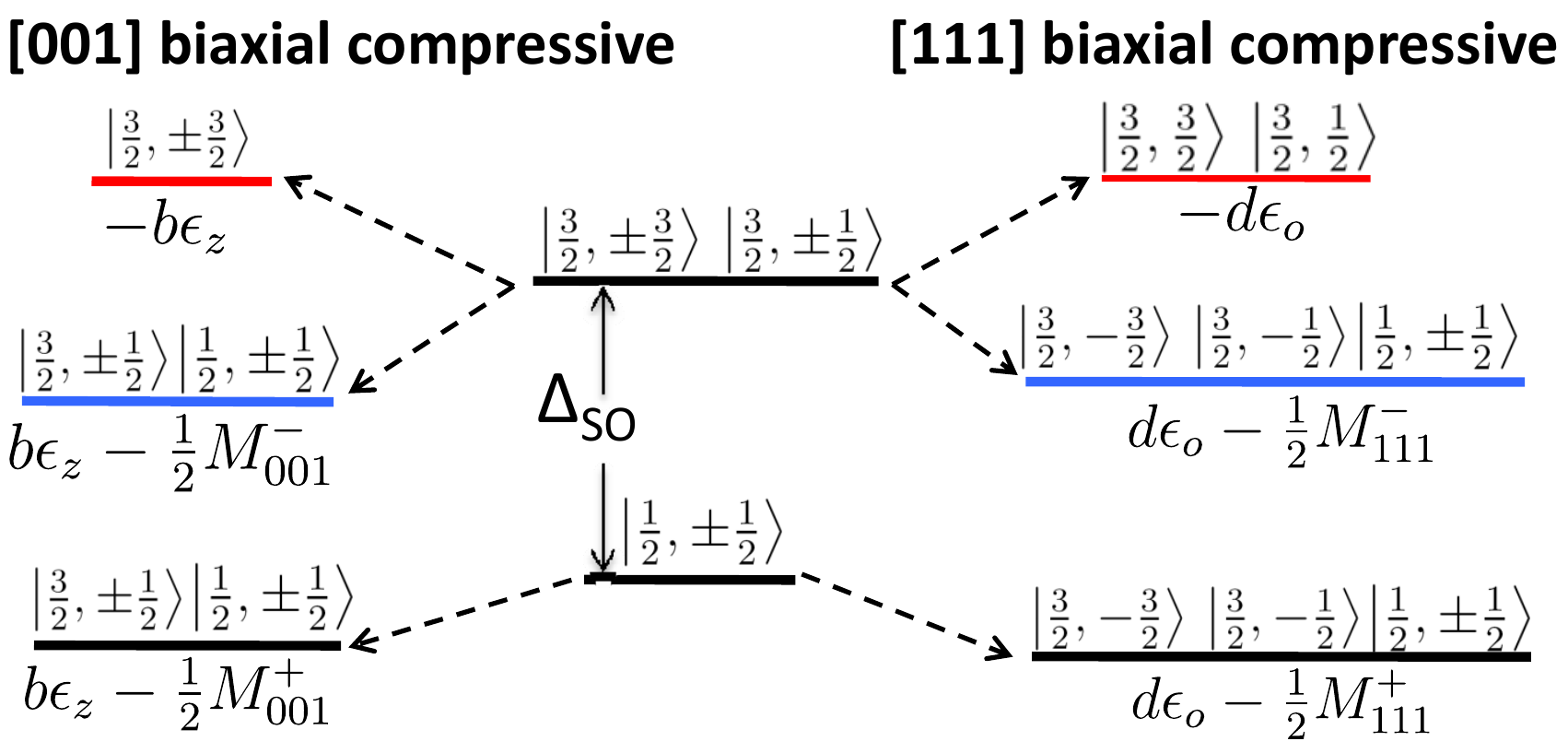}
\caption{Energy split and state mixing in the top of the valence band for two configurations of biaxial compressive strain.  In case of tensile strain the top two bands switch positions. The hole state components are denoted by $\left|J,J_{\mathbf{n}}\right\rangle$, where in the left (right) part $\mathbf{n} \parallel [001]$ ($\mathbf{n} \parallel [111]$). The mixing between hole states and energy shifts are shown next to the levels of the strained valence bands. Details about the energy parameters ($b\epsilon_z$, $d\epsilon_0$, and $M_{001}^{\pm}$ and $M_{111}^{\pm}$) are provided in Appendix~\ref{KL}.} \label{fig:strain_h_band}
\end{figure}

The effect of strain on holes is more subtle due to the HH and LH band-degeneracy in the $\Gamma$~point. The transformation properties of the 4 zone-center valence states (including spin) belong to the irreducible representation of $\Gamma_8^+$.  The amplitude of the spin-orbit coupling sets their separation from the $\Gamma_7^+$ zone-center states of the split-off valence band. By applying strain,  the $\Gamma_8^+$ energy degeneracy is lifted into two sets of spin-degenerate hole bands. Such modified bands and states can be modeled by a strain-dependent $6\times6$ Luttinger-Kohn Hamiltonian (Appendix~\ref{KL}). The left part of Fig.~\ref{fig:strain_h_band} summarizes the main results for [001]-strain and the right part for [111]-strain. When interested in the luminescence, the most relevant band for optical transitions is the topmost band since thermal holes populate this band. When interested in absorption (optical orientation), other bands also become relevant if the photon energy is larger than their energy gap from the conduction band.

\subsection{Strain-dependent selection rules}\label{sec:strain_sub}

We derive spin-dependent selection rules for the polarized luminescence from strained Si and Ge. Using time reversal and crystal space inversion symmetries, each of the valence bands is two-fold spin degenerate with states written by
\begin{eqnarray}
|\mathbf{k},\mathbf{s}_1\rangle &=& \sum_{\ell=X,Y,Z} \left[a_{\ell}(\mathbf{k}) | \ell, \uparrow \rangle  + b_{\ell}(\mathbf{k}) | \ell, \downarrow \rangle\right]e^{i\mathbf{k}\cdot \mathbf{r}}, \nonumber \\
|\mathbf{k},\mathbf{s}_2\rangle &=& \sum_{\ell=X,Y,Z} \left[-b_{\ell}^{\ast}(\mathbf{k}) | \ell, \uparrow \rangle + a_{\ell}^{\ast}(\mathbf{k}) | \ell, \downarrow \rangle\right]e^{i\mathbf{k}\cdot \mathbf{r}}.  \label{eq:hole_states}
\end{eqnarray}
The states are normalized such that $\sum_{\ell}|a_{\ell}(\mathbf{k})|^2+|b_{\ell}(\mathbf{k})|^2=1$. Using [001] and [111] biaxial strain configurations, we obtain the values of $a_{\ell}$ and $b_{\ell}$ by solving the Luttinger-Kohn Hamiltonian. To evaluate the circular polarization degree of dominant phonon-assisted optical transitions, we follow the straightforward analysis of Sec.~\ref{subsec:circ_si}, but we use Eq.~(\ref{eq:hole_states}) to represent final states in the valence band. Then we find the intensities of the right and left circularly polarized light ($I_{\pm}$) from which the circular polarization degree is extracted,
\begin{eqnarray}
P = \frac{I_{+}-I_{-}}{I_{+}+I_{+}}, \label{eq:pol_001}
\end{eqnarray}
In the biaxial strain configuration, $I_{\pm}$ depend on the in-plane strain parameter
\begin{eqnarray}
\epsilon_\parallel \equiv \frac{a_{\text{strain}}-a_{\text{relax}}}{a_{\text{relax}}},
\label{eq:in_plane_strain}
\end{eqnarray}
defined by the lattice constants of the strained and relaxed lattices.

\subsection{Silicon}\label{sec:strain_si}
We first consider the transverse phonon-assisted optical transitions which account for the dominant luminescence peaks in Si. We study the radiative recombination of spin-up electrons with general hole states in the top of the valence band [Eq.~(\ref{eq:hole_states})]. For light propagation along the [001] direction, the relative intensities of left and right circularly polarized luminescence from $x$ and $y$ valleys read
\begin{eqnarray}
I_{\pm}^{x(y)} \!\!\propto \! 1\!+\!3(|a_{X(Y)}|^2\!+\!|b_{X(Y)}|^2)\pm4\Im(a_Xa_Y^*\!-\!b_Xb_Y^*),\ \label{eq:intensity_x_y}
\end{eqnarray}
where $\Im(x)$ denotes the imaginary part of $x$. The relative intensities in the $\pm z$ valleys read
\begin{eqnarray}
I_{\pm}^z \propto 8(|a_Z|^2+|b_Z|^2). \label{eq:intensity_z}
\end{eqnarray}
The case of unstrained Si is naturally incorporated by assigning the values of $a_{\ell }$ and $b_{\ell}$ from the final states in Table \ref{Table:1}. Using these values, one can recover the intensity ratios between right and left circularly polarized luminescence in the upper panel of Fig.~\ref{fig:sr_Ge_Si}.

[001]-\textit{biaxial compressive strain ($\epsilon_\parallel < 0$)}.
The lowermost conduction valleys are along the $x$ and $y$ axes, and the topmost valence states are heavy holes (Figs.~\ref{fig:strainband} and \ref{fig:strain_h_band}). To calculate the circular polarization degree due to optical transitions of spin-up electrons, we substitute Eq.~(\ref{eq:intensity_x_y}) into Eq.~(\ref{eq:pol_001}) and assign $a_{X(Y)}=1(i)/\sqrt{2}$ and $b_X=b_Y=0$ (Table~\ref{Table:1}). The resulting circular polarization degree is
\begin{eqnarray}
P_{\Uparrow}^{001}(\epsilon_\parallel < 0) &\stackrel{\rm{TO,TA}}{\longrightarrow} & -\frac{4}{5},
\label{eq:polarization_Si_001_tensile_TO}
\end{eqnarray}
and this value is reached if the energy split of the topmost valence band is larger than $k_BT$. Due to the fact that heavy-hole states remain `pure' in this strain configuration, the $-\tfrac{4}{5}$ value is compatible with the previously derived intensity ratio of $\sigma_+$$\,$$:$$\,$$\sigma_-$$\,$=$\,$$3:27$ (see upper panel of Fig.~\ref{fig:sr_Ge_Si}).

[001]-\textit{biaxial tensile strain ($\epsilon_\parallel > 0$)}.
The lowermost conduction valleys are along the $z$ axis. The circular polarization degree of the luminescence vanishes in this case [substituting Eq.~(\ref{eq:intensity_z})  into Eq.~(\ref{eq:pol_001})],
\begin{eqnarray}
P_{\Uparrow}^{001}(\epsilon_\parallel > 0) \stackrel{\rm{TO,TA}}{\longrightarrow}  0.
\label{eq:polarization_Si_001_compressive_TO}
\end{eqnarray}
The polarization vanishes because of the symmetry of electron states in the $\pm z$ conduction valleys. It is not caused by a valence-band related effect.

[111]-\textit{biaxial strain}. Our interest switches to light propagation (and spin orientation) along $(\hat{\mathbf{x}}+\hat{\mathbf{y}}+\hat{\mathbf{z}})/\sqrt{3}$ where we assign  $\cos{\theta}=1/\sqrt{3}$ and $\phi=\pi/4$ in Eqs.~(\ref{eq:light_n})-(\ref{eq:spin_up_n}). In this strain configuration, the hole states are no longer associated with pure $\pm3/2$ or $\pm1/2$ magnetic quantum numbers (Appendix~\ref{KL}). Using analytical values of $a_{\ell}$ and $b_{\ell}$ in the topmost valence bands [Eqs.~(\ref{eq:aibi_111_compressive}) and (\ref{eq:aibi_111_tensile})], we find that the circular polarization degrees are
\begin{eqnarray}
P_{\Uparrow}^{111}(\epsilon_\parallel < 0) & \stackrel{\rm{TO,TA}}{\longrightarrow} & -\frac{1}{2}, \label{eq:polarization_Si_111_comp} \\
P_{\Uparrow}^{111}(\epsilon_\parallel > 0) & \stackrel{\rm{TO,TA}}{\longrightarrow} & \frac{1}{4}\frac{(M_{111}^+ -4 d \epsilon_{o})^2}{{M_{111}^+}^2-2M_{111}^+ d \epsilon_{o}+10d^2 \epsilon_{o}^2},\,\,\,\,\,\,\,\,\,\,\, \label{eq:polarization_Si_111_tens}
\label{eq:polarization_Si_111}
\end{eqnarray}
where $\epsilon_{o}$ relates to the off-diagonal component of the shear strain tensor,
\begin{eqnarray}
\epsilon_{o}=-\sqrt{3}\epsilon_\parallel\frac{c_{11}+2c_{12}}{c_{11}+2c_{12}+4c_{44}}. \label{eq:epsilon_0value}
\end{eqnarray}
$c_{ij}$ are the elastic stiffness constants which in silicon have values of $c_{11}=166$~GPa, $c_{12}=64$~GPa and $c_{44}=79.6$~GPa. Other parameters in Eq.~(\ref{eq:polarization_Si_111_tens}) are $d$ and $M_{111}^+$. The former is the shear deformation potential of holes associated with [111] strain. Its value is $d=-5.1$~eV in silicon. Finally,
\begin{eqnarray}
M_{111}^+ = d \epsilon_{o} +\Delta_{\text{SO}}+\sqrt{9d^2 \epsilon_{o}^2+2d \epsilon_{o}\Delta_{\text{SO}}+\Delta_{\text{SO}}^2},\,\,\,
\label{eq:eps_M}
\end{eqnarray}
relates to the strain-induced energy shift of the split-off band (see Fig.~\ref{fig:strain_h_band} for illustration). $\Delta_{\text{SO}}$$\,$=$\,$44~meV is the energy spacing between the topmost and split-off valence bands in the absence of strain.

By assigning  $\epsilon_0 \rightarrow 0$ in Eqs.~(\ref{eq:polarization_Si_111_comp}) and (\ref{eq:polarization_Si_111_tens}), we reach in agreement with the respective intensity ratios of 3$\,$:$\,$9 and 5$\,$:$\,$3 shown in the upper panel of Fig.~\ref{fig:sr_Ge_Si_111}. With increasing the strain amplitude in the tensile configuration, the coupling with split-off states becomes stronger and the circular polarization degree eventually vanishes. The smallness of $\Delta_{\text{SO}}$ in silicon is such that the drop in polarization is evident already at moderate levels of biaxial tensile strain. Quantitatively it is seen from Eq.~(\ref{eq:polarization_Si_111_tens}) by noting that $M_{111}^+ \rightarrow 4d \epsilon_{o}$ when $d \epsilon_{o} > \Delta_{\text{SO}}$.

To complete the analytical study of silicon, we repeat the procedure for the weaker LO phonon-assisted optical transitions. In this case, the relative intensities of left and right circularly polarized luminescence are governed by $\Delta_5$ intermediate states in the valence band.\cite{Li_PRL10} Using hole states from the Luttinger-Kohn Hamiltonian and considering the contribution from the lowermost conduction valley, we get
\begin{eqnarray}
&P_{\Uparrow}^{001}(\epsilon_\parallel < 0)&\!\! \stackrel{\rm{LO}}{\longrightarrow}  0,
\label{eq:polarization_Si_001_tensile_LO}\\
&P_{\Uparrow}^{001}(\epsilon_\parallel > 0)& \!\!\stackrel{\rm{LO}}{\longrightarrow}  -1,
\label{eq:polarization_Si_001_compressive_LO} \\
&P_{\Uparrow}^{111}(\epsilon_\parallel < 0)& \!\! \stackrel{\rm{LO}}{\longrightarrow}  \frac{3}{5}, \label{eq:polarization_Si_LO_111_comp} \\
P_{\Uparrow}^{111}(\epsilon_\parallel > 0) \!\!\!\!\!\!\!\! \!\!\!\!\!&\stackrel{\rm{LO}}{\longrightarrow}&\!\!\!\!\!\!\!\!\!\!\!\!\! -\frac{(M_{111}^+ -4 d \epsilon_{o})^2}{3{M_{111}^+}^2-8M_{111}^+ d \epsilon_{o}+32d^2 \epsilon_{o}^2}.\,\,\,\,\,\,\,\,\,\, \label{eq:polarization_Si_LO_111_tens}
\end{eqnarray}

\subsection{Germanium}\label{sec:strain_ge}
We first consider the LA phonon-assisted optical transitions which account for the dominant luminescence peak in Ge. As discussed in the previous section, these optical transitions are mediated by relatively pure-spin intermediate states of $\Gamma_7^-$ symmetry. Therefore, the spin angular momentum is kept during the phonon-assisted virtual transition between the $L_6^+$ and $\Gamma_7^-$ conduction-band states. This feature leads to similarity between the selection rules of direct band-gap semiconductors and those of LA phonon-assisted optical transitions in Ge.

[001]-\textit{biaxial strain}. This configuration does not break the symmetry between $L$ valleys. We repeat the straightforward analysis of Sec.~\ref{subsec:circ_si} for optical transitions of spin-up electrons but with replacing the final states by Eq.~(\ref{eq:hole_states}). For light propagation along the $z$ crystallographic axis, we get
\begin{eqnarray}
P_{\Uparrow}  \stackrel{\rm{LA}}{=} -\frac{2\Im(a_Xa_Y^*-b_Xb_Y^*)}{1 -|a_Z|^2-|b_Z|^2}.
\label{eq:polarization_ge}
\end{eqnarray}
Summing the HH and LH states in Table~\ref{Table:1}, one gets $P=50\%$  as expected from a $3:1$ intensity ratio. Also similar to direct band-gap semiconductors is that application of [001]~strain leads to $\pm$100\% circular polarization degree for optical transitions of spin-up electrons ($s_z=+1/2$). It is caused by the strain-induced energy splitting of the valence band to heavy holes that emit only $\sigma_+$ light via $J_z=+3/2$ states, or a mixture of light and split-off holes that emit only $\sigma_-$ light via $J_z=-1/2$ states.

[111]-\textit{biaxial strain}. This configuration also leads to $\pm$100\% circular polarization degree for light propagation along the [111] crystallographic axis. Since only $J_{[111]}=+3/2$ and $J_{[111]}=-1/2$ components can take part in optical transitions with spin-up electrons, a complete circular polarization is guaranteed as long as these two components are not mixed in the topmost valence band. This situation is indeed valid for [111]-biaxial strain (Appendix~\ref{KL}). The $\pm$100\% circular polarization degrees are compatible with the 3$\,$:$\,$0 and 0$\,$:$\,$1 intensity ratios shown in the lower panel of Fig.~\ref{fig:sr_Ge_Si_111}.

We discuss the less intense TO phonon-assisted optical transitions in Ge. As mentioned before, they are governed by the $L_3^-$ intermediate states. For [001] strain, by summing the contributions from all four equivalent valleys, the relative light intensity reads
\begin{eqnarray}
I_{\pm} \propto 4+3(|a_{Z}|^2+|b_{Z}|^2)\mp6\Im(a_Xa_Y^*-b_Xb_Y^*).\label{eq:intensity_TO_Ge}
\end{eqnarray}
In the compressive case, the topmost valence band includes only heavy holes [Eq.~(\ref{eq:aibi_001_compressive})] and the resulting circular polarization degree becomes
\begin{eqnarray}
P_{\Uparrow}^{001}(\epsilon_\parallel < 0)  \stackrel{\rm{TO}}{\longrightarrow}  -\frac{3}{4}. \label{eq:polarization_Ge_001_TO_comp}
\end{eqnarray}
This value is expected from the 1:7 intensity ratio of heavy holes shown in the lower panel of Fig.~\ref{fig:sr_Ge_Si}. The case of tensile strain is more involved because states of the topmost valence band are a mixture of LH and split-off states. Using Eq.~(\ref{eq:aibi_001_tensile}) and after some algebra we get
\begin{eqnarray}
P_{\Uparrow}^{001}(\epsilon_\parallel > 0)  \stackrel{\rm{TO}}{\longrightarrow}  \frac{({M_{001}^+}-4b \epsilon_{z})^2}{{6M_{001}^+}^2+8b \epsilon_{z}{M_{001}^+}+40b^2 \epsilon_{z}^2},\label{eq:polarization_Ge_001_TO_tens}\,\,\,\,\,\,\,\,
\end{eqnarray}
where $\epsilon_{z}$ relates to the diagonal components of the shear strain tensor,
\begin{eqnarray}
\epsilon_{z} = -\epsilon_\parallel\frac{c_{11}+2c_{12}}{c_{11}}. \label{eq:eps_z}
\end{eqnarray}
For Ge the elastic stiffness constants are $c_{11}=133$~GPa, $c_{12}=49$~GPa and $c_{44}=69$~GPa. Other parameters in Eq.~(\ref{eq:polarization_Ge_001_TO_tens}) are $b$ and $M_{001}^+$. The former is the shear deformation potential of holes associated with [001] strain. Its value is $b=-2.3$~eV in Ge. Finally,
\begin{eqnarray}
M_{001}^+ = b \epsilon_{z} +\Delta_{\text{SO}}+\sqrt{9b^2\epsilon_{z}^2+2b\epsilon_{z}\Delta_{\text{SO}}+\Delta_{\text{SO}}^2} \label{eq:M001Ge}
\end{eqnarray}
relates to the strain-induced energy shift of the split-off band (see Fig.~\ref{fig:strain_h_band}). By comparing Ge and Si [Eqs.~(\ref{eq:polarization_Ge_001_TO_tens})~and~(\ref{eq:polarization_Si_111_tens})], we can see that the decay to zero polarization is much slower in Ge with increasing the strain amplitude ($\Delta_{\text{SO}}$=0.044~eV in Si and 0.29~eV in Ge). Also note that for application of infinitesimal strain ($\epsilon_z\rightarrow 0$), the $\frac{1}{6}$ result by Eq.~(\ref{eq:polarization_Ge_001_TO_tens}) is compatible to the 7$\,$:$\,$5 intensity ratio of the LH states shown in the lower panel of Fig.~\ref{fig:sr_Ge_Si}.

The effect of [111]~strain on the TO phonon-assisted optical transitions in Ge is slightly more involved. It depends on hole-state mixing and also on the lowermost conduction valley due to the symmetry breaking between $L$ valleys. The circular polarization degrees become
\begin{eqnarray}
P_{\Uparrow}^{111}(\epsilon_\parallel < 0) & \stackrel{\rm{TO}}{\longrightarrow} & -\frac{9}{43}, \label{eq:polarization_Ge_111_TO_comp} \\
P_{\Uparrow}^{111}(\epsilon_\parallel > 0) & \stackrel{\rm{TO}}{\longrightarrow} & \frac{1}{2}\frac{({M_{111}^+}-4d \epsilon_{o})^2}{{M_{111}^+}^2-2d \epsilon_{o}{M_{111}^+}+10d^2 \epsilon_{o}^2}. \,\,\,\,\,\,\label{eq:polarization_Ge_111_TO_tens}
\end{eqnarray}
By assigning  $\epsilon_0 \rightarrow 0$ in Eqs.~(\ref{eq:polarization_Ge_111_TO_comp}) and (\ref{eq:polarization_Ge_111_TO_tens}), we reach in agreement with the respective intensity ratios of 34$\,$:$\,$52 and 81$\,$:$\,$27 shown in the lower panel of Fig.~\ref{fig:sr_Ge_Si_111}.

Table~\ref{Table:sr_summary} summarizes the analytical results of this and the previous sections. The listed changes in the circular polarization degrees from unstrained to strained values, $\alpha \rightarrow \beta$, are observable once the energy split between the HH and LH bands exceeds $k_BT$ (and when applicable, also once the energy split between conduction valleys is larger than $k_BT$). In some tensile cases, indicated by $\alpha \rightarrow \beta \Rightarrow 0$ for Ge and by $\alpha \rightarrow \beta \rightarrow 0$ for Si, the polarization eventually vanishes when the strain amplitude further increases. Presenting the selection rules in Si and Ge with $\Rightarrow$ and $\rightarrow$, respectively, is meant to recall that when increasing the strain amplitude the polarization drops much faster in Si because of the proximity to its split-off band.

\begin{table}
\renewcommand{\arraystretch}{1.25}
\tabcolsep=0.24 cm
\caption{
Strain-induced changes in the circular polarization degrees of optical transitions with spin-up electrons. The light propagates along the strain symmetry axis.}\label{Table:sr_summary}
\begin{tabular}{l l l l}
\hline
$\,$            &   Phonon              &     Biaxial Strain    &  Polarization                                                     \\\hline \hline
$\!\!\!$Ge      &   LA$\,\,\&\,\,$TA    & [001] compressive     &  $\,\,\,\,\tfrac{1}{2} \rightarrow 1$                             \\
$\,$            &   $\,$                & [001] tensile         &  $\,\,\,\,\tfrac{1}{2} \rightarrow -1$                            \\
$\,$            &   $\,$                & [111] compressive     &  $\,\,\,\,\tfrac{1}{2} \rightarrow 1$                             \\
$\,$            &   $\,$                & [111] tensile         &  $\,\,\,\,\tfrac{1}{2} \rightarrow -1$                            \\ \hline
$\,$            &   TO                  & [001] compressive     &  $-\tfrac{1}{5} \rightarrow -\tfrac{3}{4}$                        \\
$\,$            &   $\,$                & [001] tensile         &  $-\tfrac{1}{5} \rightarrow  \tfrac{1}{6} \Rightarrow 0$          \\
$\,$            &   $\,$                & [111] compressive     &  $-\tfrac{1}{5} \rightarrow -\tfrac{9}{43}$                       \\
$\,$            &   $\,$                & [111] tensile         &  $-\tfrac{1}{5} \rightarrow  \tfrac{1}{2} \Rightarrow 0$          \\ \hline \hline
$\!\!\!$Si      &   TA$\,\,\&\,\,$TO    & [001] compressive     &  $-\tfrac{1}{5} \rightarrow -\tfrac{4}{5}$                        \\
$\,$            &   $\,$                & [001] tensile         &  $-\tfrac{1}{5} \rightarrow 0$                                    \\
$\,$            &   $\,$                & [111] compressive     &  $-\tfrac{1}{5} \rightarrow -\tfrac{1}{2}$                        \\
$\,$            &   $\,$                & [111] tensile         &  $-\tfrac{1}{5} \rightarrow  \tfrac{1}{4} \rightarrow 0$          \\ \hline
$\,$            &   LO                  & [001] compressive     &  $\,\,\,\,\tfrac{1}{4} \rightarrow 0$                             \\
$\,$            &   $\,$                & [001] tensile         &  $\,\,\,\,\tfrac{1}{4} \rightarrow -1$                            \\
$\,$            &   $\,$                & [111] compressive     &  $\,\,\,\,\tfrac{1}{4} \rightarrow \tfrac{3}{5}$                  \\
$\,$            &   $\,$                & [111] tensile         &  $\,\,\,\,\tfrac{1}{4} \rightarrow -\tfrac{1}{3} \rightarrow 0$   \\ \hline \hline
\end{tabular}
\end{table}

\section{Numerical calculation of the spectra}\label{sec:spectra}

To verify our symmetry-based analytical discussion in previous sections, we perform an independent numerical study of the polarized luminescence spectra. The numerical method is similar to that of unstrained Si in Ref.~[\onlinecite{Li_PRL10}], and here we extend it to cases of unstrained Ge, biaxially-strained Ge, biaxially-strained Si, and relaxed ${\rm{Si}}_{1-x}{\rm{Ge}}_{x}$ alloys. The inverse effect of optical orientation following absorption of a circularly polarized light is discussed in Appendix~\ref{OO} for unstrained Si and Ge. In the next two paragraphes we briefly summarize the invoked numerical calculations.

To obtain electronic states and energies, we employ a spin-dependent local empirical psuedopotential method,\cite{Cheli_PRB76} where the strain is incorporated following the scheme of Rieger and Vogl.\cite{Rieger_PRB93} The phonon energy and displacement vectors are calculated by the adiabatic bond-charge model,\cite{Weber_PRB77} where the strain is incorporated following the scheme of Eryi\u{g}it and Herman.\cite{Eryigit_PRB96} The radiation-matter matrix elements are calculated by the electric dipole approximation,\cite{Yu_Cardona_Book} and the electron-phonon matrix elements by the rigid-ion approximation.\cite{Allen_PRB81}

In the calculation of the electron-phonon matrix elements, we have interpolated the pseudopotential with the help of the following functional form,\cite{Li_PRL10}
\begin{eqnarray}
V(q)=a_{1} \text{exp}(-a_{2}q^{4}) \times \text{sin}(a_{3}q+a_{4}).
\end{eqnarray}
To find the values of $a_i$ we make use of the well-known empirical pseudopotential form factors at $q =\sqrt{3}$, $\sqrt{8}$ and $\sqrt{11}$.\cite{Yu_Cardona_Book,Cheli_PRB76,Rieger_PRB93} To find a unique set of $a_i$ parameters, we also set the value of $V(q=\sqrt{3}/2)$ in Ge and of $V(q=1)$ in Si. The two values are chosen by fitting the ratio between intensities of different spectral peaks, $I_{LA}/I_{TO}$ in Ge and $I_{TO}/I_{TA}$ in Si, with empirical results.\cite{Haynes_JPCS59} Since the dichroic behavior is generally governed by symmetry arguments, the selection rules are indifferent to the fitting procedure (see supplemental material of Ref.~[\onlinecite{Li_PRL10}]). We have also verified this behavior for other interpolation functions.

To make a consistent study of several strained and relaxed configurations, we calculate the spectra due to radiative recombination between spin-up electrons and a Fermi-Dirac distribution of holes. The electrons are taken from the minima of the conduction band and the Fermi level of the holes is positioned in the top-edge of the valence band. This scenario is equivalent to the luminescence that results from injection of spin-polarized electrons into moderately doped \textit{p}-type samples.\cite{Asnin_SPJ76,Altukhov_JETP77} We expect these samples to provide optimal results in prospective experiments. Contrary to \textit{n}-type samples, the injection of spin-polarized electrons is not masked by a large background of unpolarized electrons. Contrary to intrinsic samples, the recombination time of minority electrons is greatly enhanced while their spin relaxation time is not severely compromised as in the case of heavily-doped \textit{p}-type samples.

\subsection{Spin-Dependent Luminescence of unstrained Ge} \label{result1}
The numerically calculated electronic band-structure and phonon dispersion curves of unstrained Ge were presented in Fig.~\ref{fig:Geband}(a) and (b), respectively. Figure~\ref{fig:Geband}(c) shows the spin-dependent luminescence of unstrained bulk Ge at 77~K. The emitted light propagates along the $+z$ axis, where the red and blue solid lines denote the relative intensities of $\sigma_+$ and $\sigma_-$ polarizations, respectively. The green dashed line shows the resulting circular polarization degree [Eq.~(\ref{eq:pol_001})]. The low-energy spectral peak around 0.75~eV is of TO phonon-assisted optical transitions. The polarization of this peak is about $-20\%$, in accord with the predicted unstrained value of $-\tfrac{1}{5}$ (Table~\ref{Table:sr_summary}). The dominant central peak around 0.76~eV is of LA phonon-assisted optical transitions and its polarization is positive and slightly higher than $30\%$. The polarization is lower than the predicted unstrained value of 50\% for the LA peak, mainly because of the compensation from the tail of TO peak with opposite polarization ($-20\%$). As will be shown below, this effect is suppressed at low temperatures when the TO and LA peaks become well resolved. The spectral peak in Fig.~\ref{fig:Geband}(c) around 0.78~eV is of TA phonon-assisted optical transitions. Compared with the thermal energy, $k_BT\approx6.7$~meV at 77~K, this peak is well separated from the other two spectral peaks. As a result, the effect from the opposite circular polarization degree of TO phonon-assisted optical transitions is diminished. This behavior explains why the simulated circular polarization increases to $40\%$ at this spectral region (close to the analytical value of $50\%$).

\subsection{Spin-Dependent Luminescence of Strained Ge} \label{result2}
\begin{figure}
\includegraphics[width=8.5 cm]{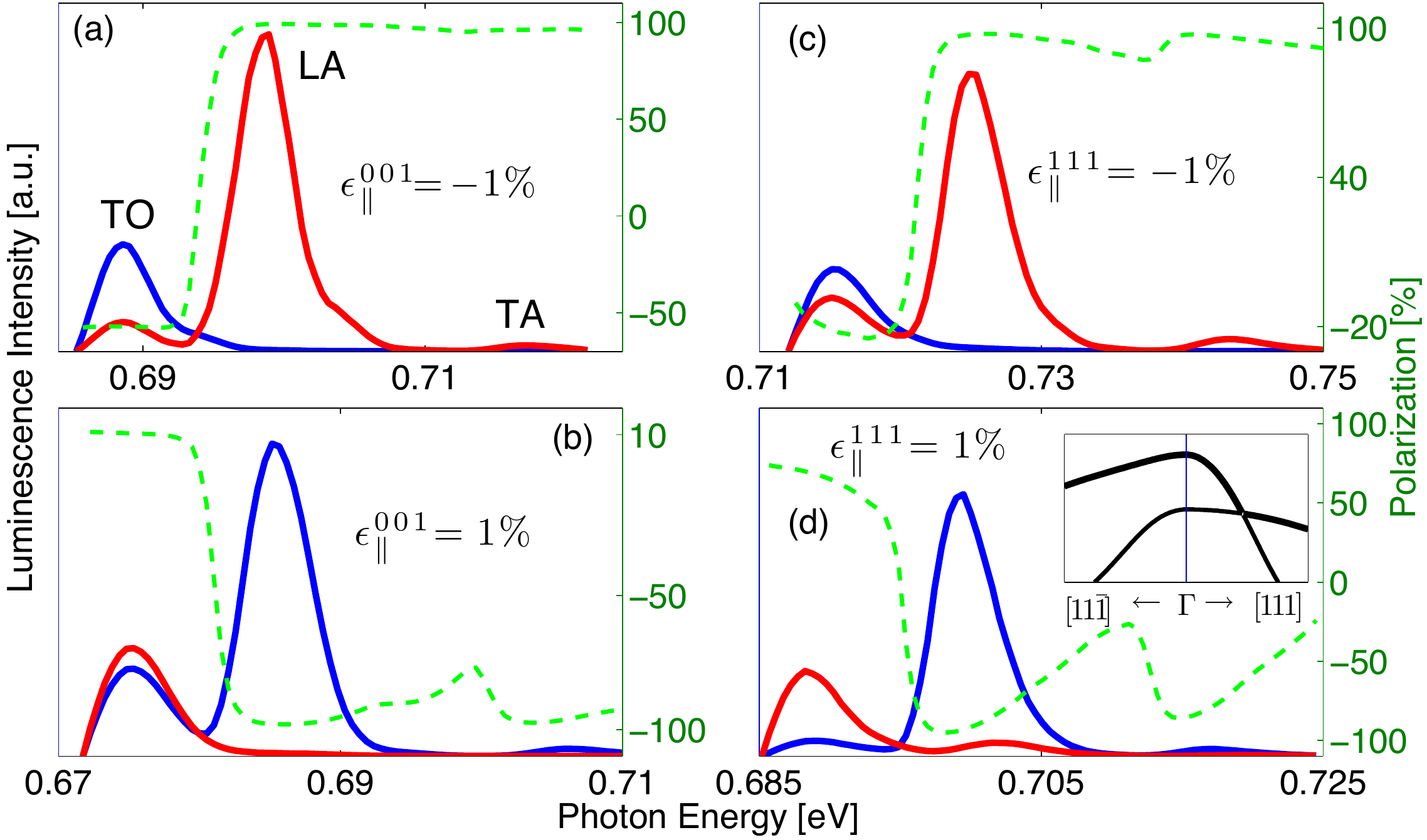}
\caption{(color online) Calculated polarized spectra of biaxially-strained germanium at 20~K. The strain configurations are [001]-compressive in (a), [001]-tensile in (b), [111]-compressive in (c),  and [111]-tensile in (d). The red (blue) solid lines denote the $\sigma_+$ ($\sigma_-$) intensities and the green dashed lines are the ensuing circular polarization degrees. The anti-crossing of the two-topmost valence bands shown in the inset of (d) indicates of the strong hole-state mixing in [111]-tensile strain.} \label{fig:strained_Ge}
\end{figure}

We simulate four configurations of $1\%$ biaxially-strained Ge. Figure~\ref{fig:strained_Ge} shows the spectra results at 20~K where the TO and LA spectral peaks are well resolved. Each of the spectra contains TO, LA and TA peaks that are red shifted compared with those in Fig.~\ref{fig:Geband}(c) due to the strain-induced energy reduction of the indirect bandgap. 

We analyze the circular polarization degrees in the spectra of strained Ge (dashed lines Fig.~\ref{fig:strained_Ge}). Consistent with the analytical picture whose results are summarized in Table~\ref{Table:sr_summary}, the dominant LA peaks show almost $\pm$100\% circular polarization degree. The one unique feature appears in the [111]-tensile configuration [Fig.~\ref{fig:strained_Ge}(d)], where the amplitude of the circular polarization degree drops from around $-$100\% in the low-energy edge of the LA peak to nearly zero in its high-energy edge. This behavior stems from the strong hole-state mixing in this strain configuration as inferred from the inset of Fig.~\ref{fig:strained_Ge}(d). This mixing increases for states away from the $\Gamma$~point so that high-energy photons can involve optical transitions with both types of holes, where each contributes oppositely to the circular polarization degree. This underlying physics explains the polarization drop in the high-energy edge tail of the LA peak in Fig.~\ref{fig:strained_Ge}(d).

Figure~\ref{fig:strained_Ge} also shows a weak peak in the high-energy part of each of the four spectra. This peak is governed by TA phonon-assisted optical transitions and its circular polarization degree is similar to that of the LA peak. As mentioned in Sec.~\ref{subsec:independent}, both LA and TA spectral features are governed by conduction-band intermediate states in the vicinity of $\Gamma_7^-$.

We conclude the discussion of the strained-Ge spectra with analysis of the TO peak. The numerically calculated circular polarization degrees for [111]-biaxial strain configurations are in agreement with the analytical picture: The polarization values in Figs.~\ref{fig:strained_Ge}(c)~and~(d) are $P_{\text{TO}} \approx -21\%$ and $P_{\text{TO}} \approx 50\%$ in accord with the predicted respective results of $-\tfrac{9}{43}$ and $\tfrac{1}{2}$ shown in Table~\ref{Table:sr_summary}. We note here that according to the analytical picture, the circular polarization degree should follow the rule $-\tfrac{1}{5} \rightarrow  \tfrac{1}{2} \Rightarrow 0$ with increasing the amplitude of the tensile strain (Table~\ref{Table:sr_summary}). However, since the split-off band in Ge is relatively well separated from the top of the valence band, the polarization does not vanishes for 1\% tensile strain (i.e., the $-\tfrac{1}{5} \rightarrow  \tfrac{1}{2}$ side of the selection rule is relevant).

When applying [001]-biaxial strain in Ge, the agreement between the analytical and numerical pictures is less successful for TO phonon-assisted optical transitions. The polarization values in Figs.~\ref{fig:strained_Ge}(a)~and~(b) are $P_{\text{TO}} \approx -57\%$ and $P_{\text{TO}} \approx 10\%$ whereas the predicted respective results are $-\tfrac{3}{4}$ and $\tfrac{1}{6}$ (Table~\ref{Table:sr_summary}). These differences are attributed to the involvement of $\Gamma_7^-$ intermediate states which are no longer symmetry-forbidden. We can understand this behavior from the transformation properties of the basis function, $(2z-x-y)/\sqrt{6}$, that represents the $L$-point TO phonons (Table \ref{Table:Phonon symmetry}). Without strain, the matrix element of the electron-phonon interaction between $L_6^+$ and $\Gamma_7^-$ states vanishes because of the opposite contributions of the $2z$ and $x+y$ components. When applying [001]-biaxial strain, the $z$ component is no longer equivalent to the $x$ and $y$ components, and the transition amplitude via $\Gamma_7^-$ becomes proportional to the strain amplitude. This transition path has a contribution similar to that of the LA phonon-assisted optical transition (50\%), leading to a diminished circular polarization degree compared with the zeroth-order predicted value of $-$75\% ($-$16.6\%) in the compressive (tensile) case.

\begin{figure}
\includegraphics[width=8.5 cm]{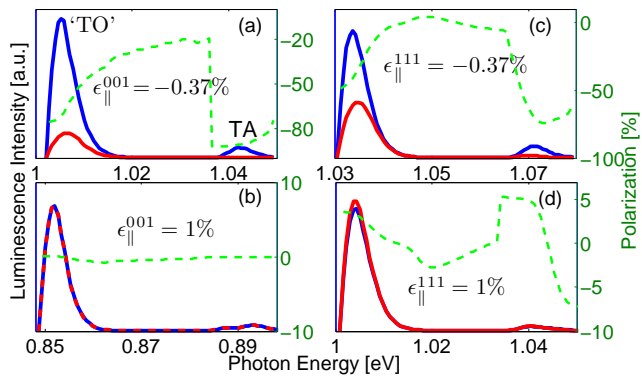}
\caption{(color online) Calculated polarized spectra of biaxially-strained silicon at 20~K. The layout is similar to that of Fig.~\ref{fig:strained_Ge}. The red and blue curves overlap in (b).} \label{fig:strained_Si}
\end{figure}

\subsection{Spin-Dependent Luminescence of Strained Si} \label{result3}
We repeat the analysis for the case of biaxially-strained Si. In the tensile configuration, we employ a stress level of $1\%$ which is accessible by growing a Si layer on a SiGe substrate. In the compressive configuration, on the other hand, there are limited substrate possibilities. Here we employ a strain level of $-$0.37\% which is accessible by growing Si on a ZnS substrate.\cite{Pfeifer_TSF72}

Figure~\ref{fig:strained_Si} shows the simulated spectra using the same layout of Fig.~\ref{fig:strained_Ge}. There are two spectral peaks in each of the four strain configurations: a weak TA peak and a dominant `TO' peak that also involves a small contribution from LO phonon-assisted optical transitions. In the compressive configurations shown in Figs.~\ref{fig:strained_Si}(a)~and~(c), the circular polarization degrees in the low-energy edge of the peaks are around $-$80\% and $-$50\%, respectively. These results are in agreement with the predicted values of $-\tfrac{4}{5}$ and $-\tfrac{1}{2}$ shown in Table~\ref{Table:sr_summary}. In both configurations, the slightly larger polarization of the TA peak compared with that of the `TO' peak stems from the opposite sign contributions of the nearly degenerate TO and LO phonon-assisted optical transitions.\cite{Li_PRL10} Figures~\ref{fig:strained_Si}(a)~and~(c) also show that the circular polarization degree drops significantly in the high-energy edges of the spectral peaks. In silicon, optical transitions that correspond to these spectral regions involve holes that are strongly susceptible to state mixing induced by the nearby split-off band.

In the biaxial tensile strain configurations shown in Figs.~\ref{fig:strained_Si}(b)~and~(d), the circular polarization degree is nearly zero across the entire spectral range. The zero-polarization is robust in the [001]-tensile configuration [Fig.~\ref{fig:strained_Si}(b)] since the electrons populate the longitudinal valleys with respect to the strain symmetry axis. Consistent with the analytical picture whose results are summarized in Table~\ref{Table:sr_summary}, TA and TO phonon-assisted optical transitions from longitudinal valleys lead to zero polarization. When applying a [111]-biaxial tensile strain, on the other hand, the energy degeneracy in the conduction band is not lifted and all six valleys are equally populated. In this configuration [Fig.~\ref{fig:strained_Si}(d)], the circular polarization degree fluctuates around a zero value because of the hole-state mixing in the valence band. According to the analytical picture, the circular polarization degree follows the rule $-\tfrac{1}{5} \rightarrow  \tfrac{1}{4} \rightarrow 0$ with increasing the strain amplitude (Table~\ref{Table:sr_summary}). The proximity of the split-off band in Si is such that the polarization nearly vanishes for 1\% tensile strain (i.e., the $\tfrac{1}{4} \rightarrow 0$ side of the selection rule is relevant). This behavior explains the 5\% circular polarization degree in the low-energy edges of the spectral peaks in Fig.~\ref{fig:strained_Si}(d).

\subsection{Spin-Polarized Luminescence from relaxed ${\rm{Si}}_{1-x}{\rm{Ge}}_{x}$ alloys}\label{result4}

Depending on the mole fractions of Si and Ge, the conduction-band edge in ${\rm{Si}}_{1-x}{\rm{Ge}}_{x}$ can be at the $L$ point or around $85\%$ along the $\Delta$ axis. The critical value of $x$ is around 0.83,\cite{Braunstein_PR_58, Weber_PRB89} at which the $L$ valleys and $\Delta$ valleys are energy degenerate. For smaller (larger) $x$ values, the polarized luminescence of the alloy is expected to exhibit Si (Ge)-like behavior.

Numerical simulations of the spin-dependent luminescence from relaxed ${\rm{Si}}_{1-x}{\rm{Ge}}_{x}$ is based on the previous procedures for Si and Ge. To obtain the band structure and electronic states, the pseudopotential Hamiltonian matrices of Si and Ge are linearly combined and solved. 
Our calculation shows a critical value at $x_c \sim 0.8325$. To obtain the phonon dispersion curve and atom displacement vectors of the alloy, the dynamical matrices of Si and Ge are also linearly combined.

\begin{figure}
\includegraphics[width=8.5 cm]{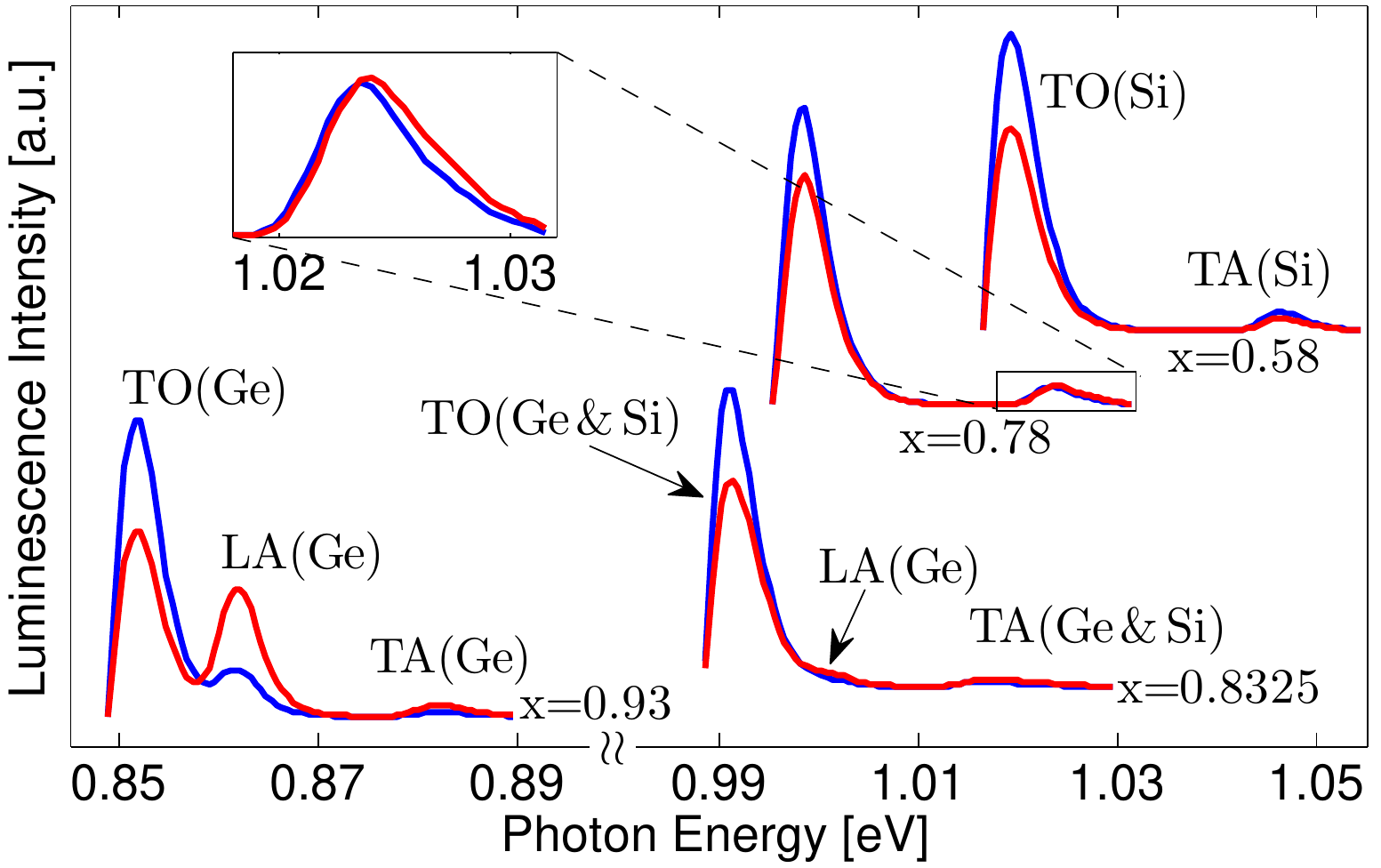}
\caption{Numerical calculation of the polarized luminescence spectra of four ${\rm{Si}}_{1-x}{\rm{Ge}}_{x}$ alloys at T=20~K. Relevant phonons are indicated next to the spectra peaks, with their origin of Si ($\Delta$ phonons) or Ge ($L$ phonons). Inset: Magnification of the TA peak for ${\rm{Si}}_{0.22}{\rm{Ge}}_{0.78}$. The crossing of $\sigma_+$ and $\sigma_-$ is due to higher order effects of optical transitions via $\Gamma_7^-$ intermediate states (see text). } \label{fig:Si_Ge}
\end{figure}

Figure~\ref{fig:Si_Ge} shows the simulated circularly-polarized luminescence due to optical transitions of spin-up electrons from four compositions of ${\rm{Si}}_{1-x}{\rm{Ge}}_{x}$ including the critical composition. As before, the simulated temperature is 20~K and the Fermi level is positioned in the top edge of the valence band. The four spectra are plotted on a single axis to demonstrate the red shift with increasing the Ge content in the alloy. A breaking at the energy axis between 0.89 and 0.99 eV indicates that such shifting is much slower when $x<x_c$ than it is when $x>x_c$. This behavior agrees with the knowledge that when $x$ decreases from 1 to 0, the $L$ valleys are raised faster than the $\Delta$ valleys.\cite{Braunstein_PR_58, Weber_PRB89}

We first discuss the Ge-like spectrum of ${\rm{Si}}_{.07}{\rm{Ge}}_{.93}$ (lower left part of Fig.~\ref{fig:Si_Ge}). Compared with Fig.~\ref{fig:Geband}(c), the most distinctive difference is the small amplitude of the LA peak. On the other hand, the circular polarization degrees of the peaks in the spectrum of ${\rm{Si}}_{.07}{\rm{Ge}}_{.93}$ follow the selection rules of bulk Ge (Fig.~\ref{fig:sr_Ge_Si}). The reason is that spin-dependent selection rules depend mostly on symmetries of the wavefunctions, whereas the amplitude of optical transitions is susceptible to the changes in the band structure. The relatively small amplitude of the LA peak is due to the increased energy gap between the $L_6^+$ and $\Gamma_7^-$ valleys [larger denominator value in Eq.~(\ref{eq:intensity})]. With reducing the Ge fraction in the alloy, this energy gap increases due to a faster shift toward higher energies of the $\Gamma_7$ conduction-band valley compared with the $L$ and $X$~valleys.

The crossing from Ge-type luminescence to Si-type luminescence in ${\rm{Si}}_{1-x}{\rm{Ge}}_{x}$ alloys takes place at relatively high mole fraction of Ge. The upper right part of Fig.~\ref{fig:Si_Ge} shows that the luminescence from ${\rm{Si}}_{0.42}{\rm{Ge}}_{0.58}$ is already Si-like in nature (other than an obvious red shift in energy). The circular polarization degrees of the TO and TA peaks in this simulated spectrum are around $-$20\% as expected from the intensity ratio of $\sigma_+$$\,$:$\,$$\sigma_-$$\,$=$\,$2$\,$:$\,$3 for these type of optical transitions in silicon.\cite{Li_PRL10} However, the Si-like behavior is gradually modified if we keep on increasing the mole fraction of Ge toward $x_c$ (in spite of the fact that electrons are still populated in $\Delta$ valleys). As an example, Fig.~\ref{fig:Si_Ge} also shows the polarized luminescence spectrum from ${\rm{Si}}_{0.22}{\rm{Ge}}_{0.78}$.  The inset shows the unique behavior of the TA peak where the $\sigma_+$ and $\sigma_-$ intensities are crossed around 1.023 eV. Here, the polarization is governed by the intensity ratio $\sigma_+$$\,$:$\,$$\sigma_-$$\,$=$\,$2$\,$:$\,$3 in the low-energy edge, while it decreases to zero and eventually flips sign in the high-energy edge. This behavior is explained by the effect of intermediate states in the vicinity of $\Gamma_7^-$. Using a $\mathbf{k}\cdot \mathbf{p}$ expansion to describe these states, we can incorporate small valence-band components with amplitudes proportional to the ratio between $k$ and the direct energy band-gap.\cite{Cardona_PR66} These components enable transitions with TA and TO phonons via intermediate states around $\Gamma_7^-$. The resulting intensity ratio of $\sigma_+$$\,$:$\,$$\sigma_-$$\,$=$\,$3$\,$:$\,$1 competes with the previous 2$\,$:$\,$3 ratio and even overcomes it when the energy of the $\Gamma_7^-$ valley is low enough (by increasing $x$).

Lastly we study the luminescence of the critical alloy ($x_c \sim 0.8325$ in our simulation), where electrons are distributed in the degenerate $L$ and $\Delta$ valleys. Figure~\ref{fig:Si_Ge} shows that this luminescence is dominated by TO phonon-assisted optical transitions whose circular polarization degree is -20\% in agreement with the derived intensity ratio of $\sigma_+:\sigma_- = 2:3$ for both Si and Ge. The luminescence also shows weak LA and TA peaks with positive circular polarization degrees of  $P_{\text{LA}}\approx$~30\% and $P_{\text{TA}}\approx$~17\%. $P_{\text{LA}}$ is lower than the theoretical value of 50\% due to the overlap of the LA peak with the tail of the dominant but oppositely-polarized TO peak. The small amplitude of the LA peak, compared with pure Ge, is attributed to a significant larger energy gap between the $L_6^+$ and $\Gamma_7^-$ valleys. $P_{\text{TA}}$ is a trade-off of the two competing effects we previously mentioned for the TA peak in ${\rm{Si}}_{0.22}{\rm{Ge}}_{0.78}$.

\section{Summary}\label{sec:sum}

We have presented a comprehensive theory of the phonon-assisted optical transitions in Si, Ge and SiGe semiconductors. The effects of [001] and [111] biaxial strains on the optical properties are studied for both compressive and tensile configurations. The changes of the circular polarization degrees from unstrained to strained-crystal values are summarized in Table~\ref{Table:sr_summary}. The concise information presented in this table together with the numerically calculated spectra presented in Figs.~\ref{fig:strained_Ge}-\ref{fig:Si_Ge} will be effective in interpretation of experiments that study the spin polarization of electrons from optical transitions in these indirect band-gap semiconductors.

The results of our analysis also elucidate two important aspects that complement the analytical picture presented in Table~\ref{Table:sr_summary}. First, in silicon the proximity  between the split-off band and the top-edge of the valence band leads to significant mixing between hole states away from the zone center. Consequently, the circular polarization degree drops rapidly from the maximal attainable value in the low-energy tail of the spectral peak to nearly zero in its high-energy tail. In addition, the significant hole-state mixing leads to quenching of the circular polarization degree when applying a large [111] biaxial tensile strain in silicon. Second, we have identified and quantified the important effect of virtual transitions with the $\Gamma_7^{-}$ intermediate states. To a large extent, changes in the circular polarization degrees of different ${\rm{Si}}_{1-x}{\rm{Ge}}_{x}$ alloys are caused by the fast energy shift of the $\Gamma_7$ conduction-band valley when varying the alloy composition. Symmetry breaking by strain also leads to changes in the circular polarization degrees by enabling virtual transitions with the $\Gamma_7^{-}$ intermediate states that are otherwise forbidden.

This work is supported by NSF Contract No.~ECCS-1231570 and by DTRA Contract No.~HDTRA1-13-1-0013.

\appendix
\section{Strain-dependent Luttinger-Kohn Hamiltonian} \label{KL}

To model the hole-state mixing as well as the energy shifts and splits in the valence band, we render a 6$\times$6 Luttinger-Kohn Hamiltonian.\cite{Chuang_98} When incorporating strain, this model is also known as the Bir-Pikus Hamiltonian.\cite{Bir_Pikus_Book} The zone-center basis functions of the valence band are chosen in the order of
\begin{widetext}
\begin{eqnarray}
\left\{\frac{X\!\!+\!iY}{\sqrt{2}}\!\!\uparrow,\,\,\,\,\frac{X\!\!+\!iY}{\sqrt{6}}\!\!\downarrow\!-\frac{2Z}{\sqrt{6}}\!\!\uparrow, \,\,\,\,-\frac{X\!\!-\!iY}{\sqrt{6}}\!\!\uparrow\!-\frac{2Z}{\sqrt{6}}\!\!\downarrow, \,\,\,\,-\frac{X\!\!-\!iY}{\sqrt{2}}\!\!\downarrow,\,\,\,\, -\frac{X\!\!+\!iY}{\sqrt{3}}\!\!\downarrow\!-\frac{Z}{\sqrt{3}}\!\!\uparrow, \,\,\,\,-\frac{X\!\!-\!iY}{\sqrt{3}}\!\!\uparrow\!+\frac{Z}{\sqrt{3}}\!\!\downarrow\right\}.
\label{eq:basis}
\end{eqnarray}

Using this basis, the Hamiltonian matrix reads
\begin{eqnarray}
H_{LK}=-\left(\begin{array}{cccccc}
P+Q & -S & R & 0 & -\frac{S}{\sqrt{2}} & \sqrt{2}R\\
-S^* & P-Q & 0 & R &-\sqrt{2}Q & \sqrt{\frac{3}{2}}S \\
R^* & 0 & P-Q & S & \sqrt{\frac{3}{2}}S^* & \sqrt{2}Q \\
0 & R^* & S^* & P+Q & -\sqrt{2}R^* & -\frac{S^*}{\sqrt{2}} \\
-\frac{S^*}{\sqrt{2}} &-\sqrt{2}Q & \sqrt{\frac{3}{2}}S & -\sqrt{2}R & P+\Delta_{\text{SO}} &0 \\
\sqrt{2}R^*  &\sqrt{\frac{3}{2}}S^* &\sqrt{2}Q & -\frac{S}{\sqrt{2}} & 0 & P+\Delta_{\text{SO}} \end{array}\right),
\label{eq:H_LK}
\end{eqnarray}
\end{widetext}
where $\Delta_{\text{SO}}$ is the split-off energy and the other parameters include wavevector components ($k_i$) and strain-tensor components ($\epsilon_{ij}$),
\begin{eqnarray}
\!\!\! P &\!\!=\!\!& \frac{\hbar^2}{2m_0}\gamma_1(k_x^2+k_y^2+k_z^2)-a_v(\epsilon_{xx}+\epsilon_{yy}+\epsilon_{zz}),\nonumber\\
\!\!\!Q &\!\!=\!\!& \frac{\hbar^2}{2m_0}\gamma_2(k_x^2+k_y^2-2k_z^2)-\frac{b}{2}(\epsilon_{xx}+\epsilon_{yy}-2\epsilon_{zz}),\nonumber\\
\!\!\!R &\!\!=\!\!& \frac{\hbar^2}{2m_0}\sqrt{3}[-\gamma_2(k_x^2-k_y^2)+2i\gamma_3 k_xk_y],\nonumber\\
&\,\,& + \frac{\sqrt{3}}{2}b(\epsilon_{xx}-\epsilon_{yy})-id \epsilon_{xy},\nonumber\\
\!\!\!S &\!\!=\!\!& \frac{\hbar^2}{2m_0}2\sqrt{3}\gamma_3(k_x-ik_y)k_z  -d(\epsilon_{xz}-i\epsilon_{yz}). \label{eq:PQRS_epsl}
\end{eqnarray}
$\{ \gamma_1,\gamma_2, \gamma_3 \}$ are the Luttinger parameters, and $a_v$ is the volume deformation potential giving rise to a uniform strain-induced energy shift of the valence bands. $b$ and $d$ are the shear deformation potentials giving rise to the energy split in the top of the valence band when applying [001] and [111] strain, respectively. The analysis of the circular polarization degree, discussed in the main text, is almost independent of the values of these well-known empirical parameters. In the infinitesimal strain picture ($\epsilon_{ij} \rightarrow 0$), the only useful information for deriving the selection rules is that the shear deformation potentials ($b$ and $d$) are negative in Si and Ge. In the finite strain picture, we also consider the proximity of the split-off valence band in Si ($\Delta_{\text{SO}}$$\,$=$\,$44~meV). This proximity leads to a measurable variation in the circular polarization degree with a moderate increase of the photon energy.

Following the discussion in the main text, the solution of the Hamiltonian in the zone-center ($k$=0) is sufficient in order to derive the strain-dependent optical selection rules. Below, we present the results for the biaxial strain configurations studied in this paper.

\subsection*{[001] biaxial strain}
This configuration is found when thin layers are grown on lattice-mismatched [001] substrates. The strain tensor is diagonal with components,
\begin{eqnarray}
\epsilon_{xx} &=& \epsilon_{yy} = \epsilon_\parallel, \,\,\,\,\,\,\,\,\,\,\, \epsilon_{zz} = -\frac{2c_{12}}{c_{11}}\epsilon_\parallel, \nonumber\\
\epsilon_{xy} &=& \epsilon_{yz} = \epsilon_{zx} = 0,
\label{eq:strain_tensor_001}
\end{eqnarray}
where $\epsilon_\parallel$ was defined in Eq.~(\ref{eq:in_plane_strain}), and $c_{ij}$ are the elastic stiffness constants with values provided after Eq.~(\ref{eq:epsilon_0value}) for Si and after Eq.~(\ref{eq:eps_z}) for Ge. The solution of Eq.~(\ref{eq:H_LK}) for zone-center states is simple and can be compactly expressed by defining the constants,
\begin{eqnarray}
M_\pm^{001} \!\!&\equiv&\!\! b \epsilon_{z} \!+\!\Delta_{\text{SO}}\!\pm\!\sqrt{9b^2 \epsilon_{z}^2\!+\!2 b \epsilon_{z}\Delta_{\text{SO}}\!+\!\Delta_{\text{SO}}^2}, \,\,\,\,\,\,\,\,\,\,\,\,\,
\label{eq:M_001}
\end{eqnarray}
where $\epsilon_{z} = \epsilon_{zz} - \tfrac{1}{2}(\epsilon_{xx}+\epsilon_{yy})$. Using these constants, the zone-center energies are
\begin{eqnarray}
\!\!\!\! E_1 &\!\!=\!\!& a_v (\epsilon_{xx}+\epsilon_{yy}+\epsilon_{zz}) - b\epsilon_{z} , \\
\!\!\!\! E_2 &\!\!=\!\!& a_v (\epsilon_{xx}+\epsilon_{yy}+\epsilon_{zz}) + b\epsilon_{z} - \frac{1}{2}M^-_{001}, \\
\!\!\!\! E_3 &\!\!=\!\!& a_v (\epsilon_{xx}+\epsilon_{yy}+\epsilon_{zz}) + b\epsilon_{z} - \frac{1}{2}M^+_{001}. \label{eq:energies_001}
\end{eqnarray}
Recalling that $b$ is negative in both Si and Ge, the energy of the topmost band is $E_1$ in biaxial compressive strain ($\epsilon_\parallel<0$) and $E_2$ in biaxial tensile strain ($\epsilon_\parallel>0$). Apart from the mutual energy shift, the left-hand side of Fig.~\ref{fig:strain_h_band} shows the relative arrangement of $E_1$, $E_2$ and $E_3$ in the compressive configuration.

The circular polarization degree of the luminescence is governed by optical transitions with holes of the topmost band (ground state). In the compressive configuration, the topmost zone-center eigenvectors are from heavy-hole states,
\begin{eqnarray}
\bm{V}^{001}_{g,\Uparrow}(\epsilon_\parallel<0)   &=& \left[ 1,0,0,0,0,0 \right]^T, \nonumber \\
\bm{V}^{001}_{g,\Downarrow}(\epsilon_\parallel<0) &=& \left[ 0,0,0,1,0,0 \right]^T. \label{eq:states_001comp}
\end{eqnarray}
In the tensile configuration, the topmost zone-center eigenvectors are a mixture of light holes and split-off holes,
\begin{eqnarray}
\!\!\!\! \bm{V}^{001}_{g,\Uparrow}(\epsilon_\parallel>0)   &\!\!=\!\!& \frac{1}{N}\left[ 0, 0,\frac{M^+_{001}}{2\sqrt{2}},0,0,-b\epsilon_{z} \right]^T, \nonumber \\
\!\!\!\!\bm{V}^{001}_{g,\Downarrow}(\epsilon_\parallel>0)  &\!\!=\!\!& \frac{1}{N}\left[ 0, \frac{M^+_{001}}{2\sqrt{2}}, 0,0,-b\epsilon_{z},0 \right]^T\!\!. \label{eq:states_001tens}
\end{eqnarray}
$N \!\!=\!\! \sqrt{\tfrac{1}{8}{M^+_{001}}^{\!\!\!2}+b^2 \epsilon_{z}^2}$ is a normalization factor. The component of the split-off holes is proportional to the strain amplitude. In the case of infinitesimal strain, we assign $M^+_{001}/N=\sqrt{8}$ and $\epsilon_{z}=0$, and get that the eigenvectors are from pure light-hole states. Using the expansion in Eq.~(\ref{eq:hole_states}), we find the coefficients from which we derive the selection rules. For example, the coefficients of the spin-up state are
\begin{eqnarray}
a_X = -\frac{1}{\sqrt{2}}, \,\,\,\,a_Y = -\frac{i}{\sqrt{2}}, \nonumber\\
a_Z = b_X = b_Y = b_Z = 0,
\label{eq:aibi_001_compressive}
\end{eqnarray}
in the [001]-compressive configuration, and
\begin{eqnarray}
a_X &=& \frac{1}{4\sqrt{3}}\frac{M^+_{001}}{N}-\frac{1}{\sqrt{3}}\frac{b \epsilon_{z}}{N}, \nonumber\\
a_Y &=& -\frac{i}{4\sqrt{3}}\frac{M^+_{001}}{N}+\frac{i}{\sqrt{3}}\frac{b \epsilon_{z}}{N}, \nonumber \\
b_Z &=& \frac{1}{2\sqrt{3}}\frac{M^+_{001}}{N}+\frac{1}{\sqrt{3}}\frac{b \epsilon_{z}}{N},\nonumber \\
a_Z &=& b_X = b_Y = 0, \,\,\,\, \label{eq:aibi_001_tensile}
\end{eqnarray}
in the [001]-tensile configuration.

\subsection*{[111] biaxial strain}
This configuration is found when thin layers are grown on lattice-mismatched [111] substrates. The components of the strain tensor are
\begin{eqnarray}
\epsilon_{xx} &=& \epsilon_{yy} = \epsilon_{zz} = \frac{4c_{44}}{c_{11}+2c_{12}+4c_{44}}\epsilon_\parallel, \nonumber\\
\epsilon_{xy} &=& \epsilon_{yz} = \epsilon_{zx} =-\frac{c_{11}+2c_{12}}{c_{11}+2c_{12}+4c_{44}}\epsilon_\parallel.
\label{eq:strain_tensor_110}
\end{eqnarray}
The solution of Eq.~(\ref{eq:H_LK}) for zone-center states can be compactly expressed by defining the constants
\begin{eqnarray}
M^\pm_{111}\!\! &=& \!\!d \epsilon_{o} \!+\!\Delta_{\text{SO}}\!\pm\!\sqrt{9d^2 \epsilon_{o}^2\!+\!2 d \epsilon_{o}\Delta_{\text{SO}}\!+\!\Delta_{\text{SO}}^2}, \,\,\,\,\,\,\,\,\,\,\,\,\,
\label{eq:M_111}
\end{eqnarray}
where $\epsilon_{o} = (\epsilon_{xy}+\epsilon_{yz}+\epsilon_{zx})/\sqrt{3}$. Using these constants, the zone-center energies are
\begin{eqnarray}
\!\!\!\! E_1 &\!\!=\!\!& a_v(\epsilon_{xx}+\epsilon_{yy}+\epsilon_{zz}) - d \epsilon_{o} , \\
\!\!\!\! E_2 &\!\!=\!\!& a_v(\epsilon_{xx}+\epsilon_{yy}+\epsilon_{zz}) + d \epsilon_{o} - \frac{1}{2}M^-_{111}, \\
\!\!\!\! E_3 &\!\!=\!\!& a_v(\epsilon_{xx}+\epsilon_{yy}+\epsilon_{zz}) + d \epsilon_{o} - \frac{1}{2}M^+_{111}. \label{eq:energies_111}
\end{eqnarray}
Recalling that $d$ is negative in both Si and Ge, the energy of the topmost band is $E_1$ in biaxial compressive strain ($\epsilon_o<0$) and $E_2$ in biaxial tensile strain ($\epsilon_o>0$). Apart from the mutual energy shift, the right-hand side of Fig.~\ref{fig:strain_h_band} shows the relative arrangement of $E_1$, $E_2$ and $E_3$ in the compressive configuration.

The order of the two-topmost bands is set by the sign of $\epsilon_\parallel$ where we recall that $d$ is negative in both Si and Ge. Apart from the mutual energy shift, $3 a_v \epsilon_d$, this information is shown in the right-hand side of Fig.~\ref{fig:strain_h_band} for the compressive configuration ($\epsilon_\parallel<0$).

The circular polarization degree of the luminescence is governed by optical transitions with holes of the topmost band (ground state). In the compressive configuration, the topmost zone-center eigenvectors are a mixture of heavy and light holes,
\begin{eqnarray}
\bm{V}^{111}_{g,\Uparrow}(\epsilon_\parallel<0)   &\!\!=\!\!&  \left[ -\frac{1+i}{\sqrt{6}}, -\frac{i}{\sqrt{2}},0,\frac{1}{\sqrt{6}},0,0 \right]^T\!\!, \nonumber \\
\bm{V}^{111}_{g,\Downarrow}(\epsilon_\parallel<0)   &\!\!=\!\!&  \left[ \frac{1}{\sqrt{6}}, 0, \frac{i}{\sqrt{2}},-\frac{1-i}{\sqrt{6}},0,0 \right]^T\!\!. \label{eq:states_111comp}
\end{eqnarray}
In the tensile configuration, the mixture is between all species of holes,
\begin{eqnarray}
\bm{V}^{111}_{g,\Uparrow}(\epsilon_\parallel>0) &\!\!=\!\!&  \frac{1}{N}\left[ \frac{i}{2\sqrt{6}}M^+_{111}, \frac{1-i}{4\sqrt{2}}M^+_{111},0,\right.\nonumber\\
&&\left.-\frac{1+i}{4\sqrt{6}}M^+_{111},0,d\epsilon_o \right]^T\!\!, \nonumber \\
\bm{V}^{111}_{g,\Downarrow}(\epsilon_\parallel>0) &\!\!=\!\!&  \frac{1}{N}\left[ -\frac{1-i}{4\sqrt{6}}M^+_{111}, 0, \frac{1+i}{4\sqrt{2}}M^+_{111},\right.\nonumber\\
&&\left.  -\frac{i}{2\sqrt{6}}M^+_{111},d\epsilon_o,0 \right]^T\!\!, \label{eq:states_111tens}
\end{eqnarray}
where the normalization factor is $N = \sqrt{\frac{1}{8}{M^+_{111}}^{\!\!\!2}+d^2 \epsilon_{o}^2}$. Due to the off-diagonal nature of the strain-tensor one cannot associate the ground states with pure $\pm3/2$ or $\pm1/2$ magnetic quantum numbers. It is true even when a vanishingly small strain is applied [assign $M^+_{111}/N=\sqrt{8}$ and $\epsilon_o=0$].

In the infinitesimal strain picture we can still classify the topmost band by its effective mass: HH in the [111]-compressive configuration and LH in the [111]-tensile configuration. To better understand the spin-dependent selection rules of these bands, it is convenient to make use of the angular-momentum eigenstates of $J_{[111]} = \mathbf{J}\cdot\mathbf{n}$, where $\mathbf{J}$ is the vector of spin-3/2 angular momentum matrices and $\mathbf{n}=(\hat{\mathbf{x}}+\hat{\mathbf{y}}+\hat{\mathbf{z}})\sqrt{3}$. From the four eigenstates, $\{ | \pm\tfrac{3}{2} \rangle_{\scriptscriptstyle{111}}, | \pm\tfrac{1}{2}\rangle_{\scriptscriptstyle{111}} \}$, we get the following information when a vanishingly small strain is applied. The composition of the spin-degenerate ground states is from $J_{[111]}=+\frac{3}{2}$ and $J_{[111]}=+\frac{1}{2}$ components in the [111]-compressive configuration (heavy holes), and from $J_{[111]}=-\frac{3}{2}$ and $J_{[111]}=-\frac{1}{2}$ components in the [111]-tensile configuration (light holes). We make use of this information when deriving the selection rules for optical transitions with spin-up electrons ($S_{[111]}=\frac{1}{2}$ in this configuration).

Finally, using the expansion in Eq.~(\ref{eq:hole_states}), we find the coefficients from which we derive the selection rules. For example, the coefficients of the spin-up state are
\begin{eqnarray}
a_X &=& \frac{1+i}{2\sqrt{3}}, \,\,\,\,\,\,\,\,\,\,\,a_Y = \frac{-1+i}{2\sqrt{3}}, \,\,\,\,\,\,\,\,\,\,\,a_Z = -\frac{i}{\sqrt{3}}, \nonumber \\
b_X &=& \frac{1+i}{2\sqrt{3}}, \,\,\,\,\,\,\,\,\,\,\,b_Y = \frac{-1-i}{2\sqrt{3}}, \,\,\,\,\,\,\,\,\,\,\,b_Z = 0,
\label{eq:aibi_111_compressive}
\end{eqnarray}
for [111]-compressive strain, and
\begin{eqnarray}
a_X &=& -\frac{i}{4\sqrt{3}}\frac{M^+_{111}}{N}+\frac{1}{\sqrt{3}}\frac{d \epsilon_{o}}{N}, \,\,\,\,b_X = -\frac{1}{4\sqrt{3}}\frac{M^+_{111}}{N},\nonumber\\
a_Y &=& \frac{1}{4\sqrt{3}}\frac{M^+_{111}}{N}-\frac{i}{\sqrt{3}}\frac{d \epsilon_{o}}{N}, \,\,\,\,\,\,\,b_Y = - \frac{1}{4\sqrt{3}}\frac{M^+_{111}}{N},\nonumber\\
a_Z &=& \frac{1-i}{4\sqrt{3}}\frac{M^+_{111}}{N}, \,\,\,\,\,\,\,\,\,\,\,\,\,\,\,\,\,\,\,\,\,\,\,\,\,\,\,\,\,\,\,\,b_Z = -\frac{1}{\sqrt{3}}\frac{d \epsilon_{o}}{N},\,\,\,\,\,\,\,\,\,\,\,\,\,
\label{eq:aibi_111_tensile}
\end{eqnarray}
for [111]-tensile strain.

\section{The effect of strain on the phonon dispersion} \label{phonon}

\begin{figure} [h]
\centering
\includegraphics[width=6.9cm]{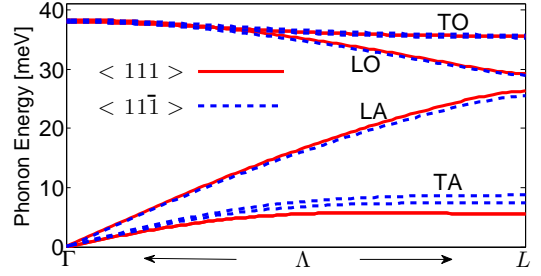}
\caption{Phonon dispersion along inequivalent $\Lambda$ symmetry axes in Ge when applying [$111$] biaxial compressive strain of 1\%. The two-fold degeneracy of transverse modes is lifted if the phonon wavevector and strain directions are nonparallel.}\label{fig:strain_phonon}
\end{figure}

As explained in the main text, changes to the magnitude, position and polarizations of the spectral peaks are mainly caused by the strain-induced degeneracy lifting and hole-state mixing.\cite{Herring_PR56,Hensel_PR65,Laude_PRB71,Fischetti_JAP96,Sun_JAP07} Changes in the phonon dispersion have a lesser effect. In the strained crystal, the bond length and bond angle between atoms change.\cite{Zi_PRB92,Sui_PRB93} However, as long as the strain levels are not very large, one can still assume an harmonic oscillation of atoms about their new equilibrium position. We have generalized the adiabatic bond-charge model to account for strain.\cite{Eryigit_PRB96} Figure~\ref{fig:strain_phonon} shows the phonon dispersion of biaxially-strained Ge in the [111] compressive configuration. From inspection of the lifted energy degeneracies, one can infer that the resulting broadening of the TA and TO spectral peaks in the luminescence of strained crystals will be negligible. Similarly, the expected energy shift of all spectral peaks is minute.

\section{Optical Orientation of Si and Ge} \label{OO}

Optical orientation is the inverse process of circularly-polarized luminescence. It involves the transfer of angular momentum from a circularly polarized light an electronic spin system.  Here we focus on the phonon-assisted optical transitions across the indirect band-gap of unstrained Si and Ge. We will not cover the absorption in the direct-gap spectral range which for the case of Ge is very effective in generating spin polarized electrons (due to the energy proximity between the $\Gamma_7$ and $L$ conduction valleys).\cite{Virgilio_PRB09,Rioux_PRB10,Guite_PRL11,Hautmann_PRB11,Loren_PRB11,Hautmann_PRB12}

\begin{figure}[h]
\includegraphics[width=8.5 cm]{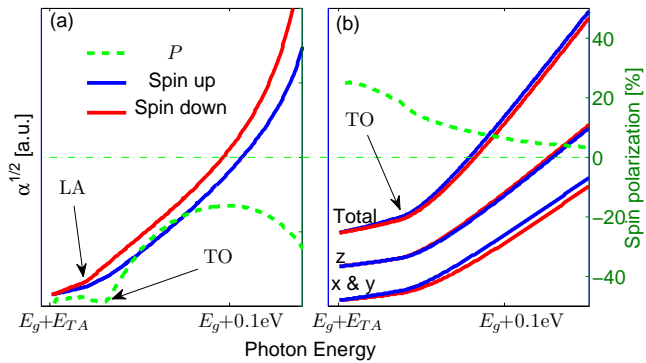}
\caption{Numerical calculation of $\alpha^{1/2}$ for spin-up and spin-down electrons by absorbing $\sigma_+$ light in Ge (a) and Si (b). Note that in Si we resolve the contributions for $\alpha^{1/2}$ from transverse and longitudinal valleys where the curves are vertically shifted for better illustration. The dashed green lines denote the spin polarization of electrons from all valleys.} \label{fig:OO}
\end{figure}

The spin polarization of photoexcited electrons is studied following absorption of a circularly-polarized light that propagates along the $+z$ axis. To interpret the simulated spin-polarization, we make use of Eq.~(\ref{eq:intensity}) and the spin-dependent selection rules in Fig.~\ref{fig:sr_Ge_Si}. Using similar numerical procedure to that of Sec.~\ref{sec:spectra}, we calculate the spin-resolved absorption coefficient $\alpha$ following light excitation with $\sigma_+$ polarization. $\alpha$ depends on the density of states in the conduction and valence bands. In bulk materials it shows quadratic dependence on the photon energy. In Fig.~\ref{fig:OO} we show the calculated value of $\alpha^{1/2}$ versus the photon energy for Si and Ge.

At low temperatures, the free-carrier absorption edge of Ge amounts to the sum of the indirect band-gap energy and the energy of the $L$$\,$-point TA phonon [Fig.~\ref{fig:OO}(a)]. Using the selection rule of this transition, the ratio between the excited spin-up and spin-down electrons should be 1$\,$:$\,$3, leading to spin polarization around $-$50\%. The onset of LA phonon-assisted optical transitions leads to enhanced absorption shown by the larger-slope region in Fig.~\ref{fig:OO}(a). Since the LA and TA selection rules are identical, the injected spin polarization of electrons does not change appreciably. The onset of TO phonon-assisted optical transitions at even higher photon energies leads to smaller spin polarization because of the opposite selection rule (3$\,$:$\,$2 versus 1$\,$:$\,$3). Finally, when the photon energy approaches the direct-gap energy (but below its absorption edge), the LA phonon-assisted optical transitions intensify and the spin polarization of electrons increases.

Figure~\ref{fig:OO}(b) shows the spin-resolved and valley-resolved value of $\alpha^{1/2}$ for Si. The absorption edge amounts to the sum of the indirect band-gap energy and the energy of TA phonons along the $\Delta$-axis near the zone-edge $X$~point. The spin polarization of electrons from all valleys is around 20\% at the absorption edge. The onset of the dominant TO phonon-assisted optical transitions leads to enhanced absorption. Most importantly, the spectral window for efficient optical spin injection in Si is of few tens of meV because of the proximity of the split-off band.

According to the selection rule in Fig.~\ref{fig:sr_Ge_Si}, the ratio between spin-up and spin-down electrons excited to $x$ or $y$ valleys in Si should be 2$\,$:$\,$1 for transverse phonon-assisted optical transitions. This information is shown in the low part of Fig.~\ref{fig:OO}(b) where the blue solid line (spin-up) is always higher than the red solid line (spin-down). For electrons excited to the $z$ valleys, the spin polarization can only come from the relatively weak LO phonon-assisted optical transitions for which the ratio is 1$\,$:$\,$3. In calculating the total spin-polarization value we have assumed that all valleys are equally populated. The spin-momentum correlations and momentum alignment during excitation were neglected.\cite{Qing_PRL11,Efanov_PSSB83} These effects may introduce quantitative but not qualitative changes (due to valley repopulation).


\begin{thebibliography}{99}
\bibitem{footnote_L3} The symmetry notations in Fig.~\ref{fig:Geband}(a) are of double-group irreducible representations with the exception of $L_3^-$ in the zone-edge of the valence band. The reason for using single-group irreducible representation in this case becomes clear in Sec.~\ref{subsec:circ_si} when we derive the selection rules of Ge.
\bibitem{footnote_nostrain} Selection rules relevant for experiments of unstrained Si and Ge are given by the summed contributions from all conduction valleys and from both types of holes. For example, in unstrained Si the sum of components in the upper panel of Fig.~\ref{fig:sr_Ge_Si} leads to the selection rules in Fig.~1(c) of Ref.~[\onlinecite{Li_PRL10}].
\end{thebibliography}
\end{document}